\documentclass{bmcart}

\usepackage[utf8]{inputenc} %unicode support
\def\includegraphics{}

\startlocaldefs
\endlocaldefs

\usepackage{hyperref}
\usepackage{psfrag}
\usepackage{chemarr}
\usepackage{graphicx}
\usepackage{amssymb}
\usepackage{amsmath}
\usepackage{xr}
\usepackage{psfrag}
\usepackage{array}
\usepackage{caption}
\usepackage{subcaption}
\begin{document}

%%% Start of article front matter
\begin{frontmatter}

\begin{fmbox}
\dochead{Research}

\title{Reliable scaling of Position Weight Matrices for binding strength comparisons between transcription factors}

%%%%%%%%%%%%%%%%%%%%%%%%%%%%%%%%%%%%%%%%%%%%%%
%%                                          %%
%% Enter the authors here                   %%
%%                                          %%
%% Specify information, if available,       %%
%% in the form:                             %%
%%   <key>={<id1>,<id2>}                    %%
%%   <key>=                                 %%
%% Comment or delete the keys which are     %%
%% not used. Repeat \author command as much %%
%% as required.                             %%
%%                                          %%
%%%%%%%%%%%%%%%%%%%%%%%%%%%%%%%%%%%%%%%%%%%%%%

\author[
   addressref={aff1,aff2},                   % id's of addresses, e.g. {aff1,aff2}
   %corref={aff1},                       % id of corresponding address, if any
   %noteref={n1},                        % id's of article notes, if any
   email={xm227@cam.ac.uk}   % email address
]{\inits{XM}\fnm{Xiaoyan} \snm{Ma}}
\author[
   addressref={aff1,aff2},
   email={de276@cam.ac.uk}
]{\inits{DE}\fnm{Daphne} \snm{Ezer}}
\author[
   addressref={aff2,aff3},
   email={cnluzon@decsai.ugr.es}
]{\inits{CN}\fnm{Carmen} \snm{Navarro}}
\author[
   addressref={aff1,aff2},                   % id's of addresses, e.g. {aff1,aff2}
   corref={aff2},                       % id of corresponding address, if any
   noteref={n1},                        % id's of article notes, if any
   email={ba255@cam.ac.uk}   % email address
]{\inits{BA}\fnm{Boris} \snm{Adryan}}

%%%%%%%%%%%%%%%%%%%%%%%%%%%%%%%%%%%%%%%%%%%%%%
%%                                          %%
%% Enter the authors' addresses here        %%
%%                                          %%
%% Repeat \address commands as much as      %%
%% required.                                %%
%%                                          %%
%%%%%%%%%%%%%%%%%%%%%%%%%%%%%%%%%%%%%%%%%%%%%%

\address[id=aff1]{%                           % unique id
  \orgname{Department of Genetics, University of Cambridge}, % university, etc
  \street{Downing Street},                     %
  \postcode{CB2 3EH}                                % post or zip code
  \city{Cambridge},                              % city
  \cny{UK}                                    % country
}
\address[id=aff2]{%
  \orgname{Cambridge Systems Biology Center, University of Cambridge},
  \street{Tennis Court Road },
  \postcode{CB2 1QR}
  \city{Cambridge},
  \cny{UK}
}
\address[id=aff3]{
  \orgname{Department of Computer Science and Artificial Intelligence, University of Granada},
  \street{Periodista Daniel Saucedo Aranda},
  %\postcode{CB2 1QR}
  \city{Granada},
  \cny{Spain}
  }
%%%%%%%%%%%%%%%%%%%%%%%%%%%%%%%%%%%%%%%%%%%%%%
%%                                          %%
%% Enter short notes here                   %%
%%                                          %%
%% Short notes will be after addresses      %%
%% on first page.                           %%
%%                                          %%
%%%%%%%%%%%%%%%%%%%%%%%%%%%%%%%%%%%%%%%%%%%%%%

\begin{artnotes}
%\note{Sample of title note}     % note to the article
\note[id=n1]{} % note, connected to author
\end{artnotes}

\end{fmbox}% comment this for two column layout

%%%%%%%%%%%%%%%%%%%%%%%%%%%%%%%%%%%%%%%%%%%%%%
%%                                          %%
%% The Abstract begins here                 %%
%%                                          %%
%% Please refer to the Instructions for     %%
%% authors on http://www.biomedcentral.com  %%
%% and include the section headings         %%
%% accordingly for your article type.       %%
%%                                          %%
%%%%%%%%%%%%%%%%%%%%%%%%%%%%%%%%%%%%%%%%%%%%%%

\begin{abstractbox}

\begin{abstract} % abstract
%\parttitle{First part title} %if any
Scoring DNA sequences against Position Weight Matrices (PWMs) is a widely adopted method to identify putative transcription factor binding sites. While common bioinformatics tools produce scores that can reflect the binding strength between a specific transcription factor and the DNA, these scores are not directly comparable between different transcription factors. Here, we provide two different ways to find the scaling parameter $\lambda$ that allows us to infer binding energy from a PWM score. The first approach uses a PWM and background genomic sequence as input to estimate $\lambda$ for a specific transcription factor, which we applied to show that $\lambda$ distributions for different transcription factor families correspond with their DNA binding properties. Our second method can reliably convert $\lambda$ between different PWMs of the same transcription factor, which allows us to directly compare PWMs that were generated by different approaches. These two approaches provide consistent and computationally efficient ways to scale PWMs scores and estimate transcription factor binding sites strength.

%\parttitle{Second part title} %if any

\end{abstract}

%%%%%%%%%%%%%%%%%%%%%%%%%%%%%%%%%%%%%%%%%%%%%%
%%                                          %%
%% The keywords begin here                  %%
%%                                          %%
%% Put each keyword in separate \kwd{}.     %%
%%                                          %%
%%%%%%%%%%%%%%%%%%%%%%%%%%%%%%%%%%%%%%%%%%%%%%

\begin{keyword}
\kwd{transcription factor}
\kwd{Position Weight Matrix}
\kwd{binding site strength}
\end{keyword}

% MSC classifications codes, if any
%\begin{keyword}[class=AMS]
%\kwd[Primary ]{}
%\kwd{}
%\kwd[; secondary ]{}
%\end{keyword}

\end{abstractbox}
%
%\end{fmbox}% uncomment this for twcolumn layout

\end{frontmatter}

%%%%%%%%%%%%%%%%%%%%%%%%%%%%%%%%%%%%%%%%%%%%%%
%%                                          %%
%% The Main Body begins here                %%
%%                                          %%
%% Please refer to the instructions for     %%
%% authors on:                              %%
%% http://www.biomedcentral.com/info/authors%%
%% and include the section headings         %%
%% accordingly for your article type.       %%
%%                                          %%
%% See the Results and Discussion section   %%
%% for details on how to create sub-sections%%
%%                                          %%
%% use \cite{...} to cite references        %%
%%  \cite{koon} and                         %%
%%  \cite{oreg,khar,zvai,xjon,schn,pond}    %%
%%  \nocite{smith,marg,hunn,advi,koha,mouse}%%
%%                                          %%
%%%%%%%%%%%%%%%%%%%%%%%%%%%%%%%%%%%%%%%%%%%%%%

%%%%%%%%%%%%%%%%%%%%%%%%% start of article main body
% <put your article body there>

%%%%%%%%%%%%%%%%
%% Background %%
%%

\section*{Content}

%%\newpage............
\section*{Introduction}
Sequence-specific transcription factors (TFs) are key elements in the regulation of gene expression. Their binding preferences to DNA have been studied extensively in vitro, in vivo and using computational methods. In vitro methods such as DNase I footprinting \cite{neph2012expansive}, protein binding microarray(PBM) \cite{mukherjee2004rapid} and high-throughput SELEX measurements \cite{roulet2002high} have provided fundamental insight into the specificity of TF binding. The systematic compilation of DNA sequences from such experiments (and along with them catalogues such as TRANSFAC \cite{matys2006transfac} or JASPAR \cite{mathelier2013jaspar}) have long suggested that TFs do not just bind to one DNA motif, but can bind to a repertoire of similar sequences. Stacks of such sequences give rise to alignment matrices, in which each column represents the absolute count of A, C, G and T nucleotide occurrences per position along the length of the motif. Work by Berg \emph{et al.}\cite{berg1987selection} introduced a derivative of the alignment or position frequency matrix (PFM), the position weight matrix (PWM), which takes the log likelihood of observing nucleotides taking their overall frequency into account. Berg \emph{et al.}\cite{berg1988selection} later showed that the score obtained by comparing the PWM against a DNA sequence is proportional to the binding energy between this TF and the DNA. In most cases the actual binding energy between the protein and DNA is not known, and the proportionality is scaled with a factor commonly termed $\lambda$.\par

There is no well-characterized and easily computable way to determine the TF binding energy to specific DNA sequences and to compare binding site strength between different types of TFs at large scale. This is problematic when scanning the genome with a library of PWMs, as scoring functions treat each PWM independently, and the absolute score associated with a ``good match" to the PWM of one transcription factor might be associated with a mismatch for another factor. A more sophisticated application of binding site strength estimation is, for example, modeling the relationship between enhancer occupancy and gene expression \cite{kim2013rearrangements,giorgetti2010noncooperative}. The experimental PBM approach \cite{mukherjee2004rapid} allows the estimation of the relative binding strength of a protein to ``naked" DNA \emph{in vitro}, but the data availability is restricted to a limited number of TFs due to high cost of the technology. In addition, PBMs are also not suitable for TFs with longer motifs, as their accuracy will decrease with the length of the DNA probe \cite{mukherjee2004rapid}. Therefore, PWM-based approaches are used to computationally estimate TF binding affinity to a specific sequence \cite{kim2013rearrangements,leith2012sequence}. \par
 
In the majority of bioinformatic studies, the scaling factor $\lambda$ is unknown and PWM scores are used at face value as measure of affinity. For example, in our own work \cite{ezer2014physical} we used the PWM score without scaling to compare binding site strength across different TFs in $\emph{E. coli}$, which might lead to a bias due to the absolute differences between the highest and lowest PWM scores across all TFs of interest. Other work has tried to assess the range of $\lambda$ on the basis of fitting calculated affinity landscapes to ChIP-seq profiles \cite{roider2007predicting,zabet2014estimating}. However, ChIP data is intrinsically noisy and the height of a ChIP peak may not accurately represent the real binding affinity, undermining the stability and accuracy of $\lambda$ obtained from these methods. In Roider \emph{et.al} \cite{roider2007predicting}, the estimated $\lambda$ for the same TF in different conditions diverged greatly in nearly one third of TFs. Furthermore, there is a wide band of possible $\lambda$ values that optimize the correlation. Aforementioned fitting methods are further reliant on chromatin accessibility data acquired under the same experimental conditions, which is often not available for specific conditions and TFs.\par

We propose a simple approximation to estimate the scaling parameter $\lambda$ based on existing PWM matrices, average maximum mismatch energy tolerance estimated by high-throughput binding energy measurements \cite{maerkl2007systems} and the distribution of PWM scores across the genome of a specific organism. This method is independent of genome-wide binding and accessibility data. Furthermore, in the cases where there are potentially inconsistent PWMs for a particular TF (e.g. derived on the basis of individual binding sites vs. derived from high-throughput efforts), we provide a method to convert the known $\lambda$ for one PWM matrix of the same TF into another suitable value for a new PWM matrix. This method is based on a computational model of the facilitated diffusion of TFs on the DNA that our group established earlier \cite{zabet2012comprehensive}. We calculate sequence-specific residence times of TFs at the DNA, which is correlated with affinity. We can therefore derive $\lambda$ for different PWMs of the same TF on the basis of the consistency of simulated residence time. These two strategies (a) calculating $\lambda$ to scale PWM scores based on the mismatch energy theory using a simple equation and (b) converting the scaling parameter $\lambda$ between different PWM matrices of the same TF on the basis of simulated residence time of facilitated diffusion \- provide simple but useful estimations of binding energy across different TFs using properly scaled PWM scores.\par

\section{Methods}
\textbf {PWM matrices for TFs for yeast, fly and vertebrates}\par
Position frequency matrices (PFM) used to construct PWM matrices were downloaded from the JASPAR database (JASPAR-CORE-2014 non-redundant PFM) \cite{mathelier2013jaspar}. Additional sources of PFM such as those contained in the BioConductor package \emph{PWMEnrich.Dmelanogaster.background} \cite{Stojnic2014PWMEnrich} were used as a source of different matrices for the same TFs. PFMs constructed with less than 30 reference sequences of validated binding sites were removed, as we deemed those insufficient descriptions of binding preference. Given that typical TF binding sites span at least six base pairs, we removed any motifs less than 6 base pairs in length.\par
A bioinformatics approach was used to derive PWM scores \cite{stormo2000dna} as follows:
\begin{equation}
S_j=\sum_{k=1}^{L}{\log_2{\frac{v_{j,k}}{f_{j+k}}}} 
\end{equation}
where j is the DNA position for the PWM score calculation, L is the length of the motif and k represents k$^{th}$ nucleotide in the PWM motif. In addition, if there is a specific nucleotide in position (j+k) on the DNA,  $f_{j+k}$ is the frequency of this nucleotide in the whole genome of a specific organism. We used a pseudo-count $\mu$ to adjust the frequency of nucleotides and obtain $v_{j,k}$ to avoid zero frequency as follows \cite{wasserman2004applied}
\begin{equation}
v_{j,k}=\frac{n_{j,k}+f_{j+k}\cdot\mu}{\sum_{x}n_{x,k}+\mu}  
\end{equation}
where $\mu$ is chosen to be 1 \cite{wasserman2004applied} and we also show that the choice of the pseudo-count $\mu$ does not have significant influence on our results (Supplementary Figure 6); $n_{x,k}$ is the frequency of certain nucleotide x in a specific position k of the motif. \par
\textbf {Simple equation to calculate $\lambda$}\par
$\lambda$ is the scaling factor that allows for direct comparison of different PWMs in terms of binding energy to DNA. The binding energy of a TF to the DNA at a specific position can be expressed as:
\begin{equation}
E=E_0 \cdot e^{-S_j/{\lambda}}
\end{equation}
where E is the binding energy, $S_j$ is the PWM score and $E_0$ is a scaling parameter. This is useful in a variety of contexts, such as comparing the binding strength of different TFs. In addition, the expected amount of time that the TF is bound to a particular DNA sequence can be estimated as:
 \begin{equation}
{\tau}_j={\tau}_0({\lambda})\cdot e^{-S_j/{\lambda}}
\end{equation}
where $S_j$ is the PWM score at position j in the genome, $\tau_0$ is the average residence time calculated as in \cite{zabet2012comprehensive}. This equation is widely used in simulations of TF binding kinetics  \cite{stormo2010determining}.\par
Given the utility of the $\lambda$ for estimating binding strength and occupancy time, it is very important to have a simple strategy for estimating it. We derive our equation based on the following core assumptions: 1. The top 0.1\% of the highest scoring matches of the PWM to intergenic regions are considered to be possible TF binding sites, as suggested by \cite{wunderlich2009different}. Their genome-wide study of different eukaryotic TFs revealed an average of 1 binding site in every 1-5 thousand base pairs of intergenic sequence. This top 0.1\% score threshold has also been similarly adopted in other studies \cite{leith2012sequence}.  2. The maximum mismatch energy between the consensus binding motif and specific DNA sequences is proportional to the information content of the PWM matrix of the TF. \par
Based on the mismatch energy theory for estimating TF binding strength \cite{berg1988selection}, the mismatch energy at a particular binding site j of TF species i in the genome can be expressed as:
\begin{equation}
E_{mismatch,i,j} ={\Delta}S_{i,j}/{\lambda_i} =(S_{max,i}-S_{i,j})/{\lambda_i} 
\end{equation}
where $S_{i,j}$ stands for the PWM score at position j, $S_{max,i}$ is for the maximum PWM score of TF species i and ${\lambda,i}$ is the scaling parameter we want to estimate. \par
The lower boundary of potential binding sites is approximated by the top 0.1\% of PWM scores following the same reason as mentioned before and corresponds to the maximum mismatch energy tolerance level as follows
\begin{equation*}
E_{maxMismatch,i}=\frac{S_{max,i}-S_{top0.1\%,i}}{\lambda_i}
\end{equation*}
where $E_{maxMismatch,i}$ stands for maximum mismatch energy tolerance for TF species i, thus, $\lambda_i$ can be calculated using:
\begin{equation}
{\lambda_i}=\frac{S_{max,i}-S_{top0.1\%,i}} {E_{maxMismatch,i}}
\end{equation}
\par
Because different transcription factors have different DNA binding domains, the maximum mismatch energy range can vary from one TF to another. Since there is only data available for 4 individual TFs using microfluidic platform-based binding energy measurements \cite{maerkl2007systems}, we estimated the maximum mismatch energy for other TFs by using the available data as the average rate and assuming that the mismatch energy tolerance is proportional to the information content of the PWM matrix as follows:
\begin{equation}
E_{maxMismatch,i}=<E_{maxMismatch}> \times \frac{If_i}{<If>}  
\end{equation}
where $If_i$ represents the information content of a specific PWM matrix, $<If>$ stands for the average information content corresponding to the average maximum mismatch energy measured by \cite{maerkl2007systems}, which is 13.2 bits. \par
We chose an average mismatch energy tolerance of 6 bits based on the study by \cite{maerkl2007systems}. They showed by mechanical trapping of molecular interactions a significant decline in binding energy by at most 2 to 3 nucleotide mismatches, and each mismatch nucleotide contributes 2 bits in mismatch energy. Even if more mutations are introduced, the binding energy does not drop further since it has already reached the background non-specific binding energy level. Another report featuring TFs from different families including: p53, Max, Glucocorticoid Receptor \cite{mueller2013quantifying} also provides additional support for 6 bits as average mismatch energy tolerance level.  \par

\textbf {Estimating $\lambda$ of a new PWM matrix for the same TF based on the residence time landscape of the facilitated diffusion model}\par
Sometimes there may be more than one PWM available for a specific TF. In order to directly compare the TF binding energy when using two alternative versions of a PWM, it is important to have a way of scaling the results by $\lambda$. $\lambda$ can be adjusted using the formalism introduced in the previous sections. As a compute-efficient alternative, we developed a more optimal strategy for estimating $\lambda$, which does not require the assumption that the PWM information content influences the energy mismatch tolerance. Instead, we base our strategy on the estimation of the sequence specific residence time of a particular TF, which is a biological meaningful quantity and can be correlated with \textit{in vitro} sequence-dependent sliding measurement of TFs \cite{leith2012sequence}. For the same TF, the sequence-specific residence time distribution calculated by Equation 4 should be as consistent as possible, even when using slightly different PWMs, if an appropriate $\lambda$ is chosen for scaling. Based on this, given a known $\lambda$ for one PWM, we are able to find another suitable $\lambda$ for the new PWM.\par

Note that the stronger the PWM score, the more likely it is that the sequence is bound by a TF and that the TF's residence time is a biologically meaningful quantity, but there is a much greater number of weak and medium strength binding sites than strong sites in the genome. Therefore, if we scored each potential binding site equally, the background of weak and medium-strength binding sites would have a greater affect on the estimated $\lambda$ than the strong binding sites. Therefore we compare residence times across different quantiles on a logarithmic binding strength scale, so that the strongest binding sites have the most influence on our $\lambda$ estimates. Specifically, in the following analysis, we take the $log_{10}$ of the cumulative distribution of PWM scores and select all binding sites with values greater than 3.0 (recall that this corresponds to the 0.1\% percent of binding sites, which were chosen as the lower boundary of weak binding sites). We divide these top-scoring binding sites into bins every 0.1 log-quantile and calculate the average residence time for each of these bins.\par

Our strategy identifies the $\lambda$ that would produce the most similar residence times for each of these log-quantiles. Assuming that for the first PWM, we already have an estimate of $\lambda$ by either binding profile fitting or other methods, we can use Equation 4 to calculate the residence time for each binding strength log-quantile, as described above. In the following analysis of this paper, since there are very few well-characterized $\lambda$ values from profile fitting, for proof-of-principle, we borrow the values obtained from Equation 6 as pre-calculated $\lambda$. Note that $\tau_0$ is calculated via the strategy described in \cite{zabet2012comprehensive} from all intergenetic regions in the genome, which has a different value for each unique PWM. \par

Now for the second PWM, we can vary $\lambda$ between the potential values of 0.1 and 3, which was shown to be a possible $\lambda$ range \cite{roider2007predicting}, and calculate the corresponding residence times at each log-quantile level. We can now compare the reference residence times from the first PWM with the residence times for the second PWM across each binding site strength level, and for each value of $\lambda$. The $\lambda$ that minimizes the mean square error between two sets of calculated residence times is chosen as the suitable $\lambda$ value for the second PWM matrix. Since outliers can have a big influence on the mean square error, we calculated the sum of the absolute differences for the natural logarithm of residence times between the two PWM matrices for these quartile bins (Equation 8) to make a comparison with the method that uses mean square error. 
\begin{equation}
\sum\nolimits_{q}|\ln{{\tau}_{q,{\lambda}}}-\ln{{\tau}_{q,ref}}|
\end{equation}
where q represents each quantile in the quantile series, ${\tau}_{q,{\lambda}}$ is the residence time in a specific quantile of a particular $\lambda$, ${\tau}_{q,ref}$ is the residence time in the same quantile of the known $\lambda$ of the reference PWM matrix. The $\lambda$ derived by these two methods show good consistency with adjusted coefficient of determination of 0.9644. Thus, there should not be significant bias using either of these two methods.\par

\section{Results}
\textbf {Estimating scaling parameter $\lambda$ for binding site affinity across different species and TF families based on Equation 6}\par
The $\lambda$ parameter is the critical link between PWM score, the estimated binding energy and TF residence time.  Estimating TF binding site affinity by comparing PWM scores at face value can lead to a large bias, especially when this includes comparisons between many types of TFs, because several properties of the PWM matrix itself can influence the PWM score. For example, the information content of the PWM matrices is positively correlated to the maximum possible PWM score, as is shown in Figure 1 with an $R^2$ value of 0.597. Thus, the absolute value of PWM scores cannot be compared directly across different TFs as an indicator of binding site strength. Therefore, proper scaling of PWM score is needed in order to compare binding site affinity across different types of TFs. Based on the methods proposed by Berg \emph{et al.} \cite{berg1988selection}, the TF binding energy for a specific binding site can be computed by Equation 3 using the estimated $\lambda$. \par

$\lambda$ calculated by this method are all within the range suggested by \cite{roider2007predicting}, which are listed in Table 1 for different organisms. The values for vertebrate species refer to all available vertebrate TFs obtained from the non-redundant PFM JASPAR database. The upper and lower bound of $\lambda$ across all organisms are quite similar, in the range of 0.25 to 2.83. This indicates that all eukaryotic TFs, no matter which organisms they belong to, all share energetically similar DNA binding mechanisms, since $\lambda$ can be interpreted as a metric for the chemical property of stickiness between the TF molecule and DNA. To demonstrate the biological application of this parameter, Figure 2 shows an example of the \emph{D. melanogaster Even-skipped stripe 1} enhancer with the comparison between PWM score and the affinity estimation using $\lambda$ scaling. The usefulness of $\lambda$ estimates becomes apparent when comparing the first two binding sites indicated by blue arrows in this locus; the second binding site has a higher PWM score, but its binding strength is lower than the first binding site once the $\lambda$ scaling factor is taken into account. Similar situations also appear in the overlapping binding site of Bicoid and Kruppel indicated by the third arrow. Thus, only comparing the raw value of PWM score \cite{ezer2014physical} may lead to false interpretations of binding site importance. When comparing binding site strength for different TFs, we need to use $\lambda$ to adjust the PWM score. \par

Next, we calculated $\lambda$ for each TF in \emph{S. cerevisiae}, \emph{D. melanogaster} and available vertebrate TFs in JASPAR \cite{mathelier2013jaspar}, which are listed in Supplemental Table 1. Figure 3 shows the overall $\lambda$ distribution in each group of organisms, showing that the mean values of the $\lambda$ distribution shift from low to high from \emph{S. cerevisiae} to \emph{D. melanogaster}, and then to the vertebrates. The difference in average information content alone does not fully explain this discrepancy (Supplement Figure 1). Interestingly, there is also a shift of mean information content from low to high values from \emph{S. cerevisiae} to vertebrates, but since the information content is the denominator in Equation 6, it cannot account as a direct cause of the observed trend in the change of $\lambda$ distribution. Instead, the range of PWM scores that fall into the top 0.1\% represented in the numerator of Equation 6 mainly contribute to the differences of the $\lambda$ distribution. \par

Furthermore, the comparison of $\lambda$ across different TF families according to the classification in JASPAR \cite{mathelier2013jaspar} is illustrated in Figure 4. Zinc-finger nuclear receptor family, the basic leucine-zipper family and helix-loop-helix family are the three families with highest average values of $\lambda$ compared with other groups with t-test p-values equal to $8.3 \cdot 10^{-5}$,$1.4 \cdot 10^{-4}$ and $6.1 \cdot 10^{-4}$ respectively. Homeobox and forkhead TF family, both of which belong to the helix-turn-helix(HTH) TF super family, appear to have lower average $\lambda$ compared with the former three families and no difference is detected between these two using t-test (p-value=0.98). The $\beta$-$\beta$-$\alpha$ zinc-finger family, the largest TF family in higher eukaryotes, shows relatively high $\lambda$ compared to the HTH super-family including the homeobox and forkhead sub-families with t-test p-value of 0.13, and it also has a wide range of $\lambda$ owing to the great diversity of binding motifs \cite{itzkovitz2006coding}. The high mobility group family shows no significant differences from HTH super-family with t-test p-value of 0.47. \par

Since $\lambda$ is the denominator to the PWM score differences between one binding site and the consensus sequence in Equation 3, a larger $\lambda$ indicates lower mismatch energy when ${\Delta}S_j$ is the same. Thus, with the same possible mismatch energy range, if $\lambda$ is larger, the PWM score can have a greater range from the consensus sequence to the potentially weakest binding site, which indicates the binding motif for the TF family has higher flexibility as suggested by \cite{Pabo1992}. This is consistent with the fact that the TFs in the zinc-finger super-family including the nuclear receptor and $\beta$-$\beta$-$\alpha$ zinc-finger families are less constrained to a particular motif than HTH super family. Additionally, cross species comparison of $\lambda$ indicates that from yeast to vertebrate, more flexible TF motifs are used, which is consistent with the result from \cite{itzkovitz2006coding} that organisms which appeared more recently in evolution tend to use more TFs with motifs of higher flexibility.\par

\textbf {Converting $\lambda$ between different PWM matrices of the same TF}\par
In many cases there are two PWMs available for the same TF, and one of these PWMs might already have a reliable estimate of $\lambda$, from any number of experimental or computational approaches \cite{zabet2014estimating}.  In such circumstances, we provide a strategy to estimate the unknown $\lambda$ associated with the alternative PWM matrix. It would be possible to calculate the unknown $\lambda$ from Equation 6, but this does not incorporate the additional data available (i.e. the known $\lambda$). Our alternative strategy not only incorporates this data, but also loosens the assumption in Equation 6 that the maximum mismatch energy for DNA binding is proportional to information content. \par

The procedure to compute a suitable $\lambda$ is based on the concept of sequence-specific residence time (Equation 4), as illustrated in Figure 5. Initially, a well-characterized $\lambda$ is computed or measured for the first PWM of a particular TF, and then we use this value to derive a $\lambda$ that is appropriate for the second PWM of the same TF.  As part of the calculation of the $\lambda$ for the second PWM, Figure 5C shows a heatmap of the estimated residence times for a TF named lame duck (lmd) in a particular binding strength quantile, at different values of $\lambda$ (ranging from 0.1 to 3.0 as suggested by both \cite{roider2007predicting} and the range of estimated $\lambda$ using Equation 6 across different organisms). Both PWMs for the TF come from FlyFactorSurvey database \cite{zhu2011flyfactorsurvey},but they are derived from different reports with motif logos shown in Figure 5B. Blank regions in the heatmap indicate the choice of $\lambda$ would generate a residence time outside the range of pre-calculated possible residence times using the first PWM and the existing $\lambda$ value, implying that these $\lambda$ values for the second PWM are unsuitable. As shown in the heatmap, blank regions often appear in very low values of $\lambda$, while if $\lambda$ is too large, the possible residence time range from weak to strong binding sites is often very restricted, which means high affinity sites cannot be distinguished from low affinity sites efficiently. $\lambda$ values with residence times all within the reference range can be further selected, as specified in Methods. Figure 5D-F compares the residence time values between two different PWMs, at different values of $\lambda$ for the second PWM. We see that the $\lambda$ in Figure 5D and 5F would not allow for consistent residence times between the two PWMs, but Figure 5E does provide consistent results. Therefore, the $\lambda$ adopted in Figure 5E is picked up as the suitable value for the second PWM matrix. More examples of residence time heatmaps for converting $\lambda$ between different PWMs are shown in Supplementary Figure 3. \par

In order to evaluate the consistency of $\lambda$ estimation between the above method and using Equation 6, we use the examples of 20 \emph{D.melanogaster} TFs with more than 1 version of PWMs available from different experiments. These PWMs are obtained from the \emph{BioConductor} R package \emph{PWMEnrich.Dmelanogaster.background} \cite{Stojnic2014PWMEnrich} and their labels are listed in Supplementary Table 2. Since there are only few $\lambda$ available from binding profile fitting, just for the purpose of illustration, the reference values of $\lambda$ were pre-calculated from Equation 6 instead. New $\lambda$ values for PWMs obtained from other experiments are computed using both methods and they show good consistency with each other with adjusted $R^2$ equals to 0.88 (Supplementary Figure 4). Converting $\lambda$ between these two PWMs in the opposite direction also show similar results (data not shown). It indicates that both methods provide consistent estimates of $\lambda$, even though they have different core assumptions.\par

\section{Discussion}
TF binding site strength estimation using PWM-based methods is essential for modelling TF-DNA interaction in functional genomics; but a proper scaling parameter is needed when using the PWM score to estimate TF binding energy. Therefore, we provide two independent methods for estimating the scaling parameter $\lambda$ in different conditions. The simple Equation 6 is widely applicable, since it only requires a PWM matrix as input, which is easy to implement compared to methods using fitting to ChIP-seq \cite{roider2007predicting,zabet2014estimating}. Our second method converts a $\lambda$ specific to one PWM into $\lambda$ for a different PWM of the same TF. It is based on the definition of sequence-specific residence time from the facilitated diffusion model of TFs on DNA \cite{zabet2012comprehensive}. This method is particularly useful for converting a previously estimated $\lambda$ into the one associated with a more up-to-date or otherwise alternative PWM matrix.\par

These two methods are consistent with one another (Supplementary Figure 4) and with previously established methods. For instance, Equation 6 can also provide very similar results compared with the estimated $\lambda$ from ChIP-seq data fitting for the 5 \emph{D. melanogaster} TFs in \cite{zabet2014estimating}, which is shown in Supplementary Figure 5. The consistent value range of $\lambda$ in different organisms calculated by this method provides additional support for the applicability of this simple equation. Moreover, the estimated distribution of $\lambda$ values for different TF families make sense in the light of motif choice for each TF families \cite{Luscombe2000}. For example, TFs in the zinc-finger TF super-family including nuclear receptor zinc-finger and $\beta$-$\beta$-$\alpha$ zinc-finger families have more flexible binding motifs, which can suit a wider range of possible binding sites than the helix-turn-helix super-family, which has a more restricted motif consensus sequence \cite{pabo1992transcription}. Inside the same super-family, because of differentiated DNA binding domains and functions, nuclear receptor zinc-finger and $\beta$-$\beta$-$\alpha$ zinc-finger families also show significant differences of $\lambda$ distribution, with t-test p-value of 0.006. However, some TF families belong to the same super-family and also share similar binding domain properties can have strong similarity in $\lambda$ distribution, e.g. homeobox family and forkhead family which both belong to the helix-turn-helix super-family. \par

There are some points that should be noted when using the simple equation method: first, it cannot be applied to very short TF motifs that are less than 6 base pairs in length. Since this method depends on calculating the difference between the top 0.1\% of PWM scores and the maximum score, if the motif is only 5 base pairs in length, the number of possible choices for sequence combination of 5 base pairs is only 1024, then the top 0.1\% of PWM scores is the top score. However, most eukaryotic motifs are more than 6 base pairs long. Eukaryotic TFs on average cover 15 bp of DNA with a core motif length of 8-15 bp \cite{kim2013rearrangements}. Thus, this limitation should not be a problem in the majority of cases. Another assumption in this method is that the mismatch energy tolerance range measured in bits is proportional to the information content of the PWM matrix. This assumption can deal with the bias from the differences in information content of most PWM matrices; however, it might not hold for PWM matrices with extremely high information content. For example, the yeast transcription factor IXR1 has an information content of 47 bits according to the PFM matrix from JASPAR \cite{mathelier2013jaspar}, which is substantially larger than the average information content of 13.2 bits. In that case, the binding energy will probably be overestimated, which leads to a lower $\lambda$, but these cases are very rare and only 7 PWM matrices in our analysis (less than 1.5\%) have information content greater than 20. The mean of information content and the mean of estimated $\lambda$ for each TF families are listed in Supplementary Table 3, and they show weak positive correlation with p-value of 0.056 (full distribution of information content across different organisms and in different TF families are shown in Supplementary Figure 3 and 4). Interestingly, the information content is actually the denominator in Equation 7 to calculate $\lambda$ which means the information content does not affect the results of $\lambda$ directly. It is the mismatch energy tolerance differences between different TF families that contributes to the variation in $\lambda$ distribution. \par

There are two limitations of this method which can potentially lead to some biases between different organisms and different TF families. One limitation is related to the calculation of mismatch energy tolerance in different groups of TF families. We apply a single cut-off threshold of top 0.1\% PWM score for weak binding sites suggested by \cite{wunderlich2009different},  but it could be possible for different TF families, different threshold should be used due to variations in their DNA binding domains. However, it is difficult to choose specific thresholds for every TF family based on the currently available data. Another limitation is the assumption that mismatch energy tolerance of a TF is proportional to the information content of the PWM matrix. It is possible that such relationship is not linear but more complicated, which is difficult to verify. Further, from the definition of information content of the PWM matrix, it sums up information content gain from each nucleotide \cite{stormo2010determining}, which implies longer motifs including more flanking base-pairs will have slightly higher information content compared to the shorter ones just with core motifs, which is an artefact of computation. However, there is no satisfactory way to deal with this problem. One possible solution is using the information content per nucleotide instead of the total information content, but this may be even more detrimental as the information content contributed by flanking sequences constitutes only a very small fraction compared to core motifs. Thus, if dividing total information content by the length of the motif, the dilution of information content can lead to even bigger biases. Another potential solution is trying to define a core motif from one PWM matrix, but this requires detailed knowledge about the TF of interest. Additionally, $\lambda$ will not be a reliable measure of biochemical stickiness of the TF to the DNA if the PWM itself is not an accurate representation of TF binding. A PWM assumes that each nucleotide position independently contributes to TF binding affinity, which may not be the case \cite{jolma2013dna,bulyk2002nucleotides}. In addition, the composition of the position frequency matrix of the PWM may contain biases due to the difficulties of attaining an unbiased validated binding site set. Nevertheless, $\lambda$ can give us insights about DNA binding properties of TFs.\par

Also, it should be pointed out that residence time in this paper refers to an estimate based on the biophysical model proposed in \cite{leith2012sequence,zabet2012comprehensive}. However, other papers report inconsistent scales of residence time according to different experimental approaches. For example, the residence time estimations obtained by Competition-ChIP methods \cite{Lickwar2012a} do not share the same order of magnitude compared to the residence times measured by FRAP or single molecular tracking \cite{mueller2008evidence,mueller2013quantifying,chen2014single}, which can probably be an artefact of experimental methods or alternatively, the range of residence time truly varies greatly across different TFs \cite{sung2014dnase}. Because the experimentally determined values are not comparable to each other, we simply adopt bioinformatics-based approaches to compute residence time. Since our method converts $\lambda$ between different PWM matrices of the same TF under the concept of residence time, it avoids to fit inconsistent experimental observations and potential variations in DNA-binding kinetics for different TFs.\par 

Although in many cases PWMs are not optimal representations of binding motifs, they have become almost universally adopted to identify potential TF binding sites. However, it is important to remember that the value of a PWM score is not directly correlated to the binding energy, but rather depends on the scaling parameter $\lambda$.  Previously, researchers either assumed that $\lambda$ has similar values across different PWMs or estimated it through computational intensive binding profile fitting methods \cite{roider2007predicting,zabet2014estimating}. Here we provide two simple strategies for estimating $\lambda$, which will let us more clearly link PWM scores with the energetics of TF binding.\par

%\bibliographystyle{apalike}
%\bibliography{merged_19.12.2014.bib} 

%%%%%%%%%%%%%%%%%%%%%%%%%%%%%%%%%%%%%%%%%%%%%%
%%                                          %%
%% Backmatter begins here                   %%
%%                                          %%
%%%%%%%%%%%%%%%%%%%%%%%%%%%%%%%%%%%%%%%%%%%%%%

\begin{backmatter}

\section*{Competing interests}
  The authors declare that they have no competing interests.

\section*{Author's contributions}
The original study was conceived by X.M, B.A.; The analysis was conducted by X.M.,D.E.,C.N.; The paper was written by X.M.,D.E.,B.A. All authors approved the final manuscript.
   
\section*{Acknowledgements}
We thank Nicolae Radu Zabet for helpful conversations and Robert Foy for editing the manuscript. \par
Funding: Chinese Scholarship Council (CSC) Scholarship (to X.M.); Marshall Scholarship (to D.E.); TIN2013-41990-R of DGICT, Madrid (to C.N); Royal Society University Research Fellowship (to B.A.)

%%  After submission of the .TEX file,                     %%
%%  you will be prompted to submit your .BBL file.         %%

% if your bibliography is in bibtex format, use those commands:
\bibliographystyle{bmc-mathphys} % Style BST file
%\bibliography{merged_19.12.2014.bib}      % Bibliography file (usually '*.bib' )

\newcommand{\BMCxmlcomment}[1]{}

\BMCxmlcomment{

<refgrp>

<bibl id="B1">
  <title><p>An expansive human regulatory lexicon encoded in transcription
  factor footprints</p></title>
  <aug>
    <au><snm>Neph</snm><fnm>S</fnm></au>
    <au><snm>Vierstra</snm><fnm>J</fnm></au>
    <au><snm>Stergachis</snm><fnm>AB</fnm></au>
    <au><snm>Reynolds</snm><fnm>AP</fnm></au>
    <au><snm>Haugen</snm><fnm>E</fnm></au>
    <au><snm>Vernot</snm><fnm>B</fnm></au>
    <au><snm>Thurman</snm><fnm>RE</fnm></au>
    <au><snm>John</snm><fnm>S</fnm></au>
    <au><snm>Sandstrom</snm><fnm>R</fnm></au>
    <au><snm>Johnson</snm><fnm>AK</fnm></au>
    <au><cnm>others</cnm></au>
  </aug>
  <source>Nature</source>
  <publisher>Nature Publishing Group</publisher>
  <pubdate>2012</pubdate>
  <volume>489</volume>
  <issue>7414</issue>
  <fpage>83</fpage>
  <lpage>-90</lpage>
</bibl>

<bibl id="B2">
  <title><p>Rapid analysis of the DNA-binding specificities of transcription
  factors with DNA microarrays</p></title>
  <aug>
    <au><snm>Mukherjee</snm><fnm>S</fnm></au>
    <au><snm>Berger</snm><fnm>MF</fnm></au>
    <au><snm>Jona</snm><fnm>G</fnm></au>
    <au><snm>Wang</snm><fnm>XS</fnm></au>
    <au><snm>Muzzey</snm><fnm>D</fnm></au>
    <au><snm>Snyder</snm><fnm>M</fnm></au>
    <au><snm>Young</snm><fnm>RA</fnm></au>
    <au><snm>Bulyk</snm><fnm>ML</fnm></au>
  </aug>
  <source>Nature Genetics</source>
  <publisher>Nature Publishing Group</publisher>
  <pubdate>2004</pubdate>
  <volume>36</volume>
  <issue>12</issue>
  <fpage>1331</fpage>
  <lpage>-1339</lpage>
</bibl>

<bibl id="B3">
  <title><p>High-throughput SELEX--SAGE method for quantitative modeling of
  transcription-factor binding sites</p></title>
  <aug>
    <au><snm>Roulet</snm><fnm>E</fnm></au>
    <au><snm>Busso</snm><fnm>S</fnm></au>
    <au><snm>Camargo</snm><fnm>AA</fnm></au>
    <au><snm>Simpson</snm><fnm>AJ</fnm></au>
    <au><snm>Mermod</snm><fnm>N</fnm></au>
    <au><snm>Bucher</snm><fnm>P</fnm></au>
  </aug>
  <source>Nature Biotechnology</source>
  <publisher>Nature Publishing Group</publisher>
  <pubdate>2002</pubdate>
  <volume>20</volume>
  <issue>8</issue>
  <fpage>831</fpage>
  <lpage>-835</lpage>
</bibl>

<bibl id="B4">
  <title><p>TRANSFAC and its module TRANSCompel: transcriptional gene
  regulation in eukaryotes</p></title>
  <aug>
    <au><snm>Matys</snm><fnm>V</fnm></au>
    <au><snm>Kel Margoulis</snm><fnm>OV</fnm></au>
    <au><snm>Fricke</snm><fnm>E</fnm></au>
    <au><snm>Liebich</snm><fnm>I</fnm></au>
    <au><snm>Land</snm><fnm>S</fnm></au>
    <au><snm>Barre Dirrie</snm><fnm>A</fnm></au>
    <au><snm>Reuter</snm><fnm>I</fnm></au>
    <au><snm>Chekmenev</snm><fnm>D</fnm></au>
    <au><snm>Krull</snm><fnm>M</fnm></au>
    <au><snm>Hornischer</snm><fnm>K</fnm></au>
    <au><cnm>others</cnm></au>
  </aug>
  <source>Nucleic Acids Research</source>
  <publisher>Oxford Univ Press</publisher>
  <pubdate>2006</pubdate>
  <fpage>D108</fpage>
  <lpage>-D110</lpage>
</bibl>

<bibl id="B5">
  <title><p>JASPAR 2014: an extensively expanded and updated open-access
  database of transcription factor binding profiles</p></title>
  <aug>
    <au><snm>Mathelier</snm><fnm>A</fnm></au>
    <au><snm>Zhao</snm><fnm>X</fnm></au>
    <au><snm>Zhang</snm><fnm>AW</fnm></au>
    <au><snm>Parcy</snm><fnm>F</fnm></au>
    <au><snm>Worsley Hunt</snm><fnm>R</fnm></au>
    <au><snm>Arenillas</snm><fnm>DJ</fnm></au>
    <au><snm>Buchman</snm><fnm>S</fnm></au>
    <au><snm>Chen</snm><fnm>Cy</fnm></au>
    <au><snm>Chou</snm><fnm>A</fnm></au>
    <au><snm>Ienasescu</snm><fnm>H</fnm></au>
    <au><cnm>others</cnm></au>
  </aug>
  <source>Nucleic Acids Research</source>
  <publisher>Oxford Univ Press</publisher>
  <pubdate>2013</pubdate>
  <fpage>gkt997</fpage>
</bibl>

<bibl id="B6">
  <title><p>Selection of DNA Binding Sites by Regulatory Proteins
  Statistical-mechanical Theory and Application to Operators and
  Promoters</p></title>
  <aug>
    <au><snm>Berg</snm><fnm>OG</fnm></au>
    <au><snm>Hippel</snm><fnm>PH</fnm></au>
  </aug>
  <source>J. Mol. Biol</source>
  <pubdate>1987</pubdate>
  <volume>193</volume>
  <fpage>723</fpage>
  <lpage>-750</lpage>
</bibl>

<bibl id="B7">
  <title><p>Selection of DNA binding sites by regulatory proteins</p></title>
  <aug>
    <au><snm>Berg</snm><fnm>OG</fnm></au>
    <au><snm>Hippel</snm><fnm>PH</fnm></au>
  </aug>
  <source>Trends in Biochemical Sciences</source>
  <publisher>Elsevier</publisher>
  <pubdate>1988</pubdate>
  <volume>13</volume>
  <issue>6</issue>
  <fpage>207</fpage>
  <lpage>-211</lpage>
</bibl>

<bibl id="B8">
  <title><p>Rearrangements of 2.5 kilobases of noncoding DNA from the
  Drosophila even-skipped locus define predictive rules of genomic
  cis-regulatory logic</p></title>
  <aug>
    <au><snm>Kim</snm><fnm>AR</fnm></au>
    <au><snm>Martinez</snm><fnm>C</fnm></au>
    <au><snm>Ionides</snm><fnm>J</fnm></au>
    <au><snm>Ramos</snm><fnm>AF</fnm></au>
    <au><snm>Ludwig</snm><fnm>MZ</fnm></au>
    <au><snm>Ogawa</snm><fnm>N</fnm></au>
    <au><snm>Sharp</snm><fnm>DH</fnm></au>
    <au><snm>Reinitz</snm><fnm>J</fnm></au>
  </aug>
  <source>PLoS Genetics</source>
  <publisher>Public Library of Science</publisher>
  <pubdate>2013</pubdate>
  <volume>9</volume>
  <issue>2</issue>
  <fpage>e1003243</fpage>
</bibl>

<bibl id="B9">
  <title><p>Noncooperative interactions between transcription factors and
  clustered DNA binding sites enable graded transcriptional responses to
  environmental inputs</p></title>
  <aug>
    <au><snm>Giorgetti</snm><fnm>L</fnm></au>
    <au><snm>Siggers</snm><fnm>T</fnm></au>
    <au><snm>Tiana</snm><fnm>G</fnm></au>
    <au><snm>Caprara</snm><fnm>G</fnm></au>
    <au><snm>Notarbartolo</snm><fnm>S</fnm></au>
    <au><snm>Corona</snm><fnm>T</fnm></au>
    <au><snm>Pasparakis</snm><fnm>M</fnm></au>
    <au><snm>Milani</snm><fnm>P</fnm></au>
    <au><snm>Bulyk</snm><fnm>ML</fnm></au>
    <au><snm>Natoli</snm><fnm>G</fnm></au>
  </aug>
  <source>Molecular Cell</source>
  <publisher>Elsevier</publisher>
  <pubdate>2010</pubdate>
  <volume>37</volume>
  <issue>3</issue>
  <fpage>418</fpage>
  <lpage>-428</lpage>
</bibl>

<bibl id="B10">
  <title><p>Sequence-dependent sliding kinetics of p53</p></title>
  <aug>
    <au><snm>Leith</snm><fnm>JS</fnm></au>
    <au><snm>Tafvizi</snm><fnm>A</fnm></au>
    <au><snm>Huang</snm><fnm>F</fnm></au>
    <au><snm>Uspal</snm><fnm>WE</fnm></au>
    <au><snm>Doyle</snm><fnm>PS</fnm></au>
    <au><snm>Fersht</snm><fnm>AR</fnm></au>
    <au><snm>Mirny</snm><fnm>LA</fnm></au>
    <au><snm>Oijen</snm><fnm>AM</fnm></au>
  </aug>
  <source>Proceedings of the National Academy of Sciences</source>
  <publisher>National Acad Sciences</publisher>
  <pubdate>2012</pubdate>
  <volume>109</volume>
  <issue>41</issue>
  <fpage>16552</fpage>
  <lpage>-16557</lpage>
</bibl>

<bibl id="B11">
  <title><p>Physical constraints determine the logic of bacterial promoter
  architectures</p></title>
  <aug>
    <au><snm>Ezer</snm><fnm>D</fnm></au>
    <au><snm>Zabet</snm><fnm>NR</fnm></au>
    <au><snm>Adryan</snm><fnm>B</fnm></au>
  </aug>
  <source>Nucleic Acids Research</source>
  <publisher>Oxford Univ Press</publisher>
  <pubdate>2014</pubdate>
  <fpage>gku078</fpage>
</bibl>

<bibl id="B12">
  <title><p>Predicting transcription factor affinities to DNA from a
  biophysical model</p></title>
  <aug>
    <au><snm>Roider</snm><fnm>HG</fnm></au>
    <au><snm>Kanhere</snm><fnm>A</fnm></au>
    <au><snm>Manke</snm><fnm>T</fnm></au>
    <au><snm>Vingron</snm><fnm>M</fnm></au>
  </aug>
  <source>Bioinformatics</source>
  <publisher>Oxford Univ Press</publisher>
  <pubdate>2007</pubdate>
  <volume>23</volume>
  <issue>2</issue>
  <fpage>134</fpage>
  <lpage>-141</lpage>
</bibl>

<bibl id="B13">
  <title><p>Estimating binding properties of transcription factors from
  genome-wide binding profiles</p></title>
  <aug>
    <au><snm>Zabet</snm><fnm>NR</fnm></au>
    <au><snm>Adryan</snm><fnm>B</fnm></au>
  </aug>
  <source>Nucleic Acids Research</source>
  <publisher>Oxford Univ Press</publisher>
  <pubdate>2014</pubdate>
  <fpage>gku1269</fpage>
</bibl>

<bibl id="B14">
  <title><p>A systems approach to measuring the binding energy landscapes of
  transcription factors</p></title>
  <aug>
    <au><snm>Maerkl</snm><fnm>SJ</fnm></au>
    <au><snm>Quake</snm><fnm>SR</fnm></au>
  </aug>
  <source>Science</source>
  <publisher>American Association for the Advancement of Science</publisher>
  <pubdate>2007</pubdate>
  <volume>315</volume>
  <issue>5809</issue>
  <fpage>233</fpage>
  <lpage>-237</lpage>
</bibl>

<bibl id="B15">
  <title><p>A comprehensive computational model of facilitated diffusion in
  prokaryotes</p></title>
  <aug>
    <au><snm>Zabet</snm><fnm>NR</fnm></au>
    <au><snm>Adryan</snm><fnm>B</fnm></au>
  </aug>
  <source>Bioinformatics</source>
  <publisher>Oxford Univ Press</publisher>
  <pubdate>2012</pubdate>
  <volume>28</volume>
  <issue>11</issue>
  <fpage>1517</fpage>
  <lpage>-1524</lpage>
</bibl>

<bibl id="B16">
  <title><p>PWMEnrich: PWM enrichment analysis. R package version
  4.2.0.</p></title>
  <aug>
    <au><snm>Stojnic</snm><fnm>R</fnm></au>
    <au><snm>Diez</snm><fnm>D</fnm></au>
  </aug>
  <pubdate>2014</pubdate>
</bibl>

<bibl id="B17">
  <title><p>DNA binding sites: representation and discovery</p></title>
  <aug>
    <au><snm>Stormo</snm><fnm>GD</fnm></au>
  </aug>
  <source>Bioinformatics</source>
  <publisher>Oxford Univ Press</publisher>
  <pubdate>2000</pubdate>
  <volume>16</volume>
  <issue>1</issue>
  <fpage>16</fpage>
  <lpage>-23</lpage>
</bibl>

<bibl id="B18">
  <title><p>Applied bioinformatics for the identification of regulatory
  elements</p></title>
  <aug>
    <au><snm>Wasserman</snm><fnm>WW</fnm></au>
    <au><snm>Sandelin</snm><fnm>A</fnm></au>
  </aug>
  <source>Nature Reviews Genetics</source>
  <publisher>Nature Publishing Group</publisher>
  <pubdate>2004</pubdate>
  <volume>5</volume>
  <issue>4</issue>
  <fpage>276</fpage>
  <lpage>-287</lpage>
</bibl>

<bibl id="B19">
  <title><p>Determining the specificity of protein--DNA
  interactions</p></title>
  <aug>
    <au><snm>Stormo</snm><fnm>GD</fnm></au>
    <au><snm>Zhao</snm><fnm>Y</fnm></au>
  </aug>
  <source>Nature Reviews Genetics</source>
  <publisher>Nature Publishing Group</publisher>
  <pubdate>2010</pubdate>
  <volume>11</volume>
  <issue>11</issue>
  <fpage>751</fpage>
  <lpage>-760</lpage>
</bibl>

<bibl id="B20">
  <title><p>Different gene regulation strategies revealed by analysis of
  binding motifs</p></title>
  <aug>
    <au><snm>Wunderlich</snm><fnm>Z</fnm></au>
    <au><snm>Mirny</snm><fnm>LA</fnm></au>
  </aug>
  <source>Trends in Genetics</source>
  <publisher>Elsevier</publisher>
  <pubdate>2009</pubdate>
  <volume>25</volume>
  <issue>10</issue>
  <fpage>434</fpage>
  <lpage>-440</lpage>
</bibl>

<bibl id="B21">
  <title><p>Quantifying transcription factor kinetics: At work or at
  play?</p></title>
  <aug>
    <au><snm>Mueller</snm><fnm>F</fnm></au>
    <au><snm>Stasevich</snm><fnm>TJ</fnm></au>
    <au><snm>Mazza</snm><fnm>D</fnm></au>
    <au><snm>McNally</snm><fnm>JG</fnm></au>
  </aug>
  <source>Critical reviews in biochemistry and molecular biology</source>
  <publisher>Informa Healthcare USA, Inc. New York</publisher>
  <pubdate>2013</pubdate>
  <volume>48</volume>
  <issue>5</issue>
  <fpage>492</fpage>
  <lpage>-514</lpage>
</bibl>

<bibl id="B22">
  <title><p>Coding limits on the number of transcription factors</p></title>
  <aug>
    <au><snm>Itzkovitz</snm><fnm>S</fnm></au>
    <au><snm>Tlusty</snm><fnm>T</fnm></au>
    <au><snm>Alon</snm><fnm>U</fnm></au>
  </aug>
  <source>BMC Genomics</source>
  <publisher>BioMed Central Ltd</publisher>
  <pubdate>2006</pubdate>
  <volume>7</volume>
  <issue>1</issue>
  <fpage>239</fpage>
</bibl>

<bibl id="B23">
  <title><p>Transcription factors: structural families and principles of DNA
  recognition</p></title>
  <aug>
    <au><snm>Pabo</snm><fnm>CO</fnm></au>
    <au><snm>Sauer</snm><fnm>RT</fnm></au>
  </aug>
  <source>Annual Review of Biochemistry</source>
  <publisher>Annual Reviews 4139 El Camino Way, PO Box 10139, Palo Alto, CA
  94303-0139, USA</publisher>
  <pubdate>1992</pubdate>
  <volume>61</volume>
  <issue>1</issue>
  <fpage>1053</fpage>
  <lpage>-1095</lpage>
</bibl>

<bibl id="B24">
  <title><p>FlyFactorSurvey: a database of Drosophila transcription factor
  binding specificities determined using the bacterial one-hybrid
  system</p></title>
  <aug>
    <au><snm>Zhu</snm><fnm>LJ</fnm></au>
    <au><snm>Christensen</snm><fnm>RG</fnm></au>
    <au><snm>Kazemian</snm><fnm>M</fnm></au>
    <au><snm>Hull</snm><fnm>CJ</fnm></au>
    <au><snm>Enuameh</snm><fnm>MS</fnm></au>
    <au><snm>Basciotta</snm><fnm>MD</fnm></au>
    <au><snm>Brasefield</snm><fnm>JA</fnm></au>
    <au><snm>Zhu</snm><fnm>C</fnm></au>
    <au><snm>Asriyan</snm><fnm>Y</fnm></au>
    <au><snm>Lapointe</snm><fnm>DS</fnm></au>
    <au><cnm>others</cnm></au>
  </aug>
  <source>Nucleic Acids Research</source>
  <publisher>Oxford Univ Press</publisher>
  <pubdate>2011</pubdate>
  <volume>39</volume>
  <issue>suppl 1</issue>
  <fpage>D111</fpage>
  <lpage>-D117</lpage>
</bibl>

<bibl id="B25">
  <title><p>An overview of the structures of protein-DNA complexes</p></title>
  <aug>
    <au><snm>Luscombe</snm><fnm>NM</fnm></au>
    <au><snm>Austin</snm><fnm>SE</fnm></au>
    <au><snm>Berman</snm><fnm>HM</fnm></au>
    <au><snm>Thornton</snm><fnm>JM</fnm></au>
  </aug>
  <source>Genome Biol</source>
  <pubdate>2000</pubdate>
  <volume>1</volume>
  <issue>1</issue>
  <fpage>1</fpage>
  <lpage>-37</lpage>
</bibl>

<bibl id="B26">
  <title><p>Transcription factors: structural families and principles of DNA
  recognition</p></title>
  <aug>
    <au><snm>Pabo</snm><fnm>CO</fnm></au>
    <au><snm>Sauer</snm><fnm>RT</fnm></au>
  </aug>
  <source>Annual Review of Biochemistry</source>
  <publisher>Annual Reviews 4139 El Camino Way, PO Box 10139, Palo Alto, CA
  94303-0139, USA</publisher>
  <pubdate>1992</pubdate>
  <volume>61</volume>
  <issue>1</issue>
  <fpage>1053</fpage>
  <lpage>-1095</lpage>
</bibl>

<bibl id="B27">
  <title><p>DNA-binding specificities of human transcription
  factors</p></title>
  <aug>
    <au><snm>Jolma</snm><fnm>A</fnm></au>
    <au><snm>Yan</snm><fnm>J</fnm></au>
    <au><snm>Whitington</snm><fnm>T</fnm></au>
    <au><snm>Toivonen</snm><fnm>J</fnm></au>
    <au><snm>Nitta</snm><fnm>KR</fnm></au>
    <au><snm>Rastas</snm><fnm>P</fnm></au>
    <au><snm>Morgunova</snm><fnm>E</fnm></au>
    <au><snm>Enge</snm><fnm>M</fnm></au>
    <au><snm>Taipale</snm><fnm>M</fnm></au>
    <au><snm>Wei</snm><fnm>G</fnm></au>
    <au><cnm>others</cnm></au>
  </aug>
  <source>Cell</source>
  <publisher>Elsevier</publisher>
  <pubdate>2013</pubdate>
  <volume>152</volume>
  <issue>1</issue>
  <fpage>327</fpage>
  <lpage>-339</lpage>
</bibl>

<bibl id="B28">
  <title><p>Nucleotides of transcription factor binding sites exert
  interdependent effects on the binding affinities of transcription
  factors</p></title>
  <aug>
    <au><snm>Bulyk</snm><fnm>ML</fnm></au>
    <au><snm>Johnson</snm><fnm>PL</fnm></au>
    <au><snm>Church</snm><fnm>GM</fnm></au>
  </aug>
  <source>Nucleic Acids Research</source>
  <publisher>Oxford Univ Press</publisher>
  <pubdate>2002</pubdate>
  <volume>30</volume>
  <issue>5</issue>
  <fpage>1255</fpage>
  <lpage>-1261</lpage>
</bibl>

<bibl id="B29">
  <title><p>Genome-wide protein-DNA binding dynamics suggest a molecular clutch
  for transcription factor function</p></title>
  <aug>
    <au><snm>Lickwar</snm><fnm>CR</fnm></au>
    <au><snm>Mueller</snm><fnm>F</fnm></au>
    <au><snm>Hanlon</snm><fnm>SE</fnm></au>
    <au><snm>McNally</snm><fnm>JG</fnm></au>
    <au><snm>Lieb</snm><fnm>JD</fnm></au>
  </aug>
  <source>Nature</source>
  <publisher>Nature Publishing Group</publisher>
  <pubdate>2012</pubdate>
  <volume>484</volume>
  <issue>7393</issue>
  <fpage>251</fpage>
  <lpage>-255</lpage>
</bibl>

<bibl id="B30">
  <title><p>Evidence for a common mode of transcription factor interaction with
  chromatin as revealed by improved quantitative fluorescence recovery after
  photobleaching</p></title>
  <aug>
    <au><snm>Mueller</snm><fnm>F</fnm></au>
    <au><snm>Wach</snm><fnm>P</fnm></au>
    <au><snm>McNally</snm><fnm>JG</fnm></au>
  </aug>
  <source>Biophysical journal</source>
  <publisher>Elsevier</publisher>
  <pubdate>2008</pubdate>
  <volume>94</volume>
  <issue>8</issue>
  <fpage>3323</fpage>
  <lpage>-3339</lpage>
</bibl>

<bibl id="B31">
  <title><p>Single-molecule dynamics of enhanceosome assembly in embryonic stem
  cells</p></title>
  <aug>
    <au><snm>Chen</snm><fnm>J</fnm></au>
    <au><snm>Zhang</snm><fnm>Z</fnm></au>
    <au><snm>Li</snm><fnm>L</fnm></au>
    <au><snm>Chen</snm><fnm>BC</fnm></au>
    <au><snm>Revyakin</snm><fnm>A</fnm></au>
    <au><snm>Hajj</snm><fnm>B</fnm></au>
    <au><snm>Legant</snm><fnm>W</fnm></au>
    <au><snm>Dahan</snm><fnm>M</fnm></au>
    <au><snm>Lionnet</snm><fnm>T</fnm></au>
    <au><snm>Betzig</snm><fnm>E</fnm></au>
    <au><cnm>others</cnm></au>
  </aug>
  <source>Cell</source>
  <publisher>Elsevier</publisher>
  <pubdate>2014</pubdate>
  <volume>156</volume>
  <issue>6</issue>
  <fpage>1274</fpage>
  <lpage>-1285</lpage>
</bibl>

<bibl id="B32">
  <title><p>DNase footprint signatures are dictated by factor dynamics and DNA
  sequence</p></title>
  <aug>
    <au><snm>Sung</snm><fnm>MH</fnm></au>
    <au><snm>Guertin</snm><fnm>MJ</fnm></au>
    <au><snm>Baek</snm><fnm>S</fnm></au>
    <au><snm>Hager</snm><fnm>GL</fnm></au>
  </aug>
  <source>Molecular Cell</source>
  <publisher>Elsevier</publisher>
  <pubdate>2014</pubdate>
  <volume>56</volume>
  <issue>2</issue>
  <fpage>275</fpage>
  <lpage>-285</lpage>
</bibl>

</refgrp>
} % end of \BMCxmlcomment

% or include bibliography directly:
% \begin{thebibliography}
% \bibitem{b1}
% \end{thebibliography}

%%%%%%%%%%%%%%%%%%%%%%%%%%%%%%%%%%%
%%                               %%
%% Figures                       %%
%%                               %%
%% NB: this is for captions and  %%
%% Titles. All graphics must be  %%
%% submitted separately and NOT  %%
%% included in the Tex document  %%
%%                               %%
%%%%%%%%%%%%%%%%%%%%%%%%%%%%%%%%%%%

%%
%% Do not use \listoffigures as most will included as separate files

\section*{Figures}

\begin{figure}[h!]

\begin{center}

\includegraphics[scale = 0.5]{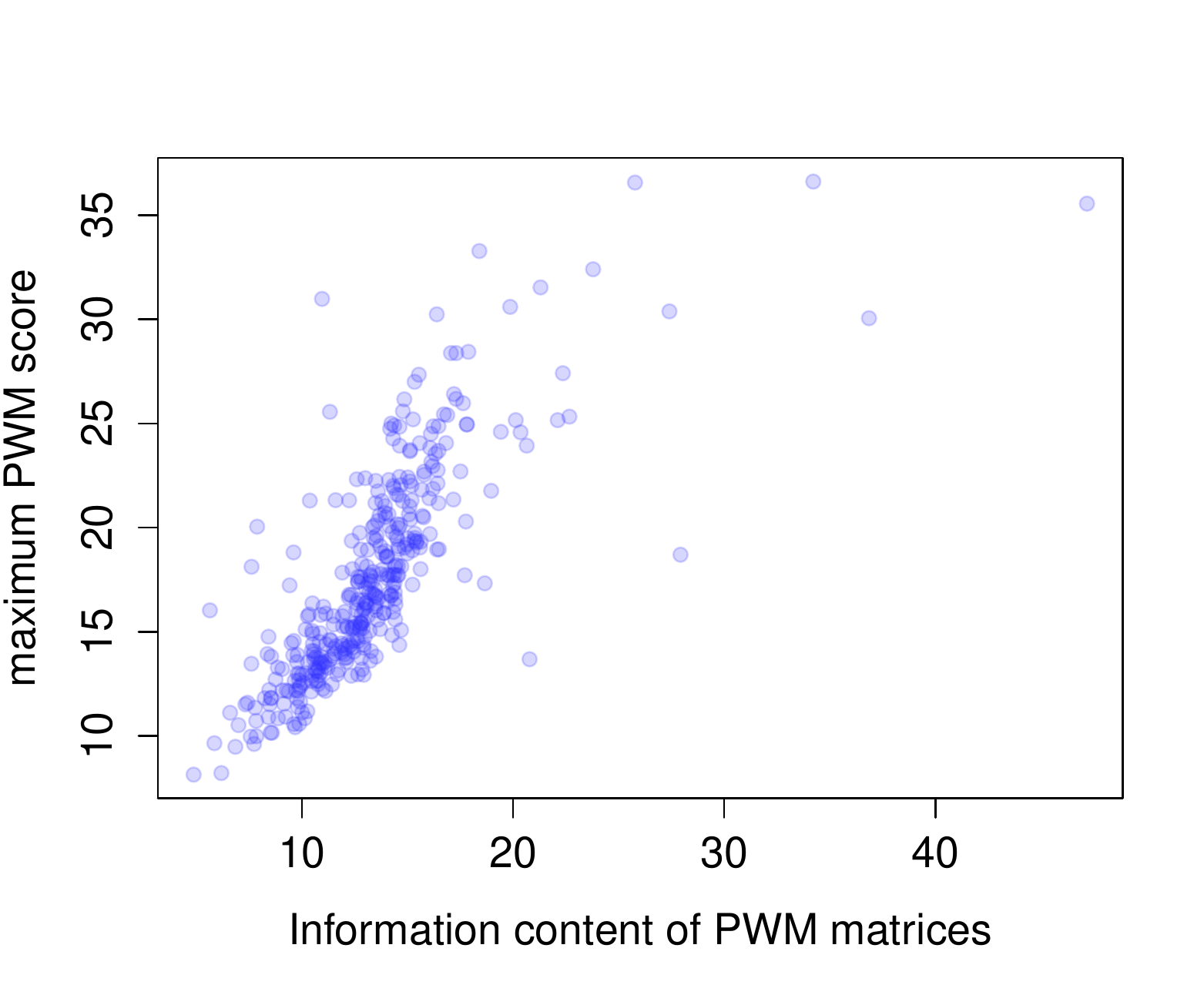}

\caption{\emph{The relationship between maximum PWM score and information content of PWM matrices.} Individual dots represents each PWM matrix generated from the non-redundant PFM JASPAR-CORE database \cite{mathelier2013jaspar} after the filtering procedures specified in the Methods section. There is a strong positive correlation between the information content of the PWM and the maximum possible PWM score that could be generated by that PWM, with an $R^2$ value of 0.597.}

\end{center}

\end{figure}

\begin{figure}[h!]

\begin{center}

\includegraphics[scale=0.33]{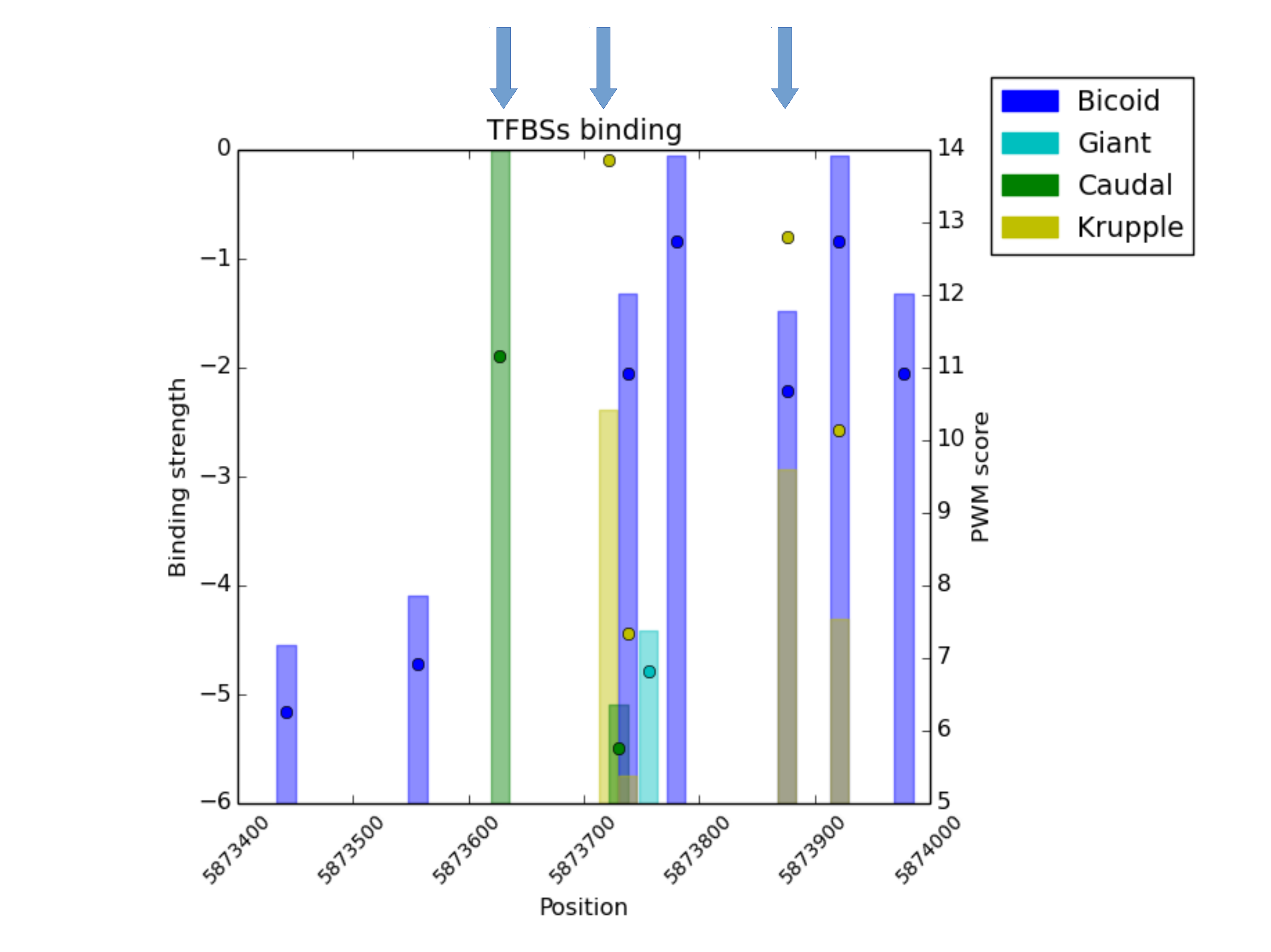}

\caption{\emph{A comparison between PWM score and binding site strength in the \emph{D. melanogaster even-skipped stripe 1 enhancer.}} The \emph{even-skipped stripe 1} enhancer on chromosome 2R is dense with binding sites. We compare the raw PWM scores (circles) and the $\lambda$-scaled binding strength (height of the bars) for each of these binding sites, colour-coded by the type of TF. Based on raw PWM scores, one might assume that the Caudal site indicated by the first blue arrow would have a lower binding strength than the Kruppel site indicated by the second blue arrow; however, once the binding strength is scaled by $\lambda$ using Equation 5, it becomes evident that the opposite is the more likely scenario. The third arrow points to a location where a Kruppel and a Bicoid binding site overlap. Here, the $\lambda$ adjusted binding strength estimates would suggest that Bicoid binding site is stronger, while a raw PWM score would suggest the opposite. These results illustrate how using raw PWM scores may result in biased interpretation of the relative binding strength of TFs.}

\end{center}

\end{figure}

\begin{figure}[h!]

\begin{center}

\includegraphics[width = 12cm]{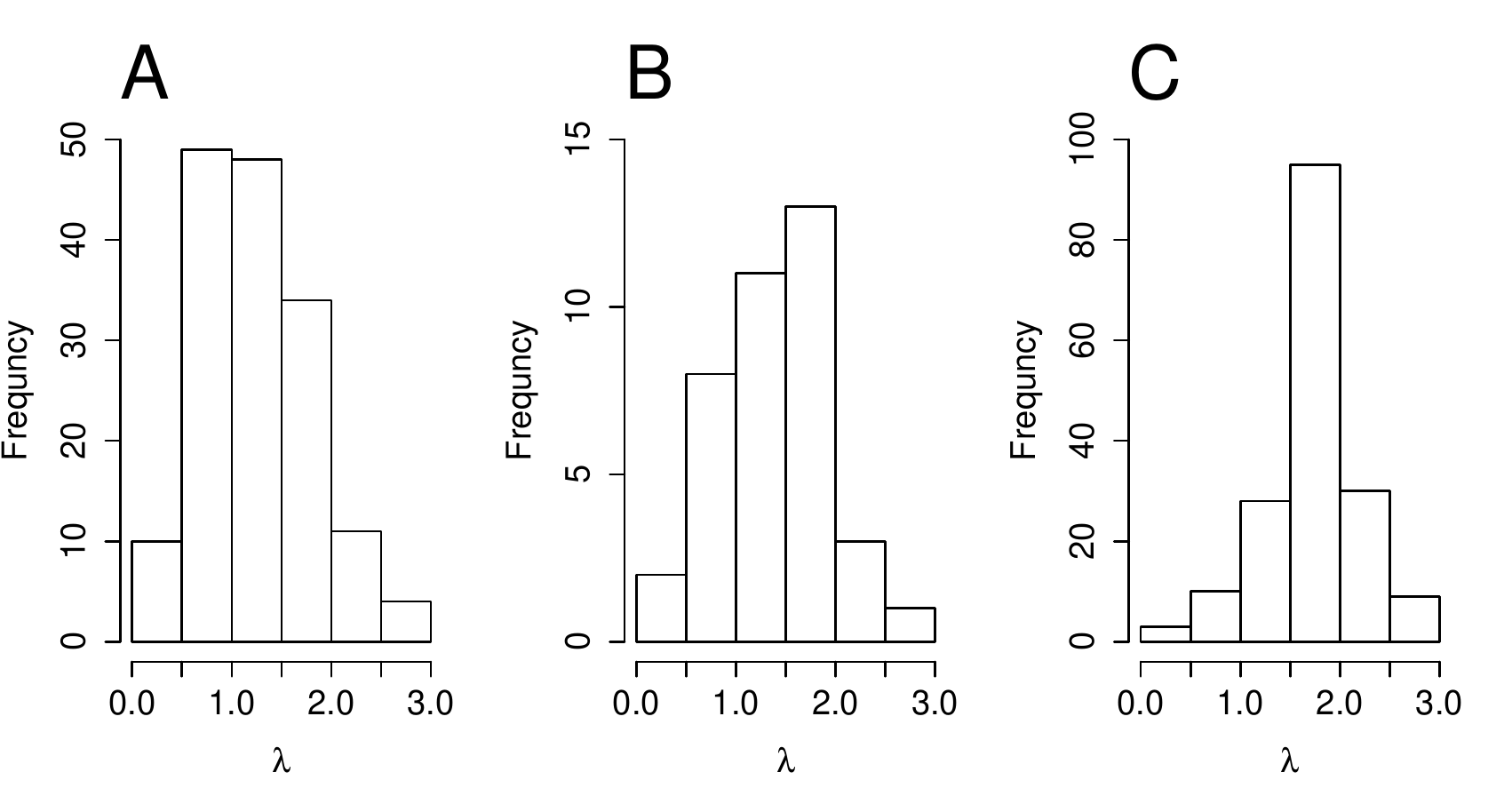}

\caption{\emph{$\lambda$ distributions across difference organisms.} The histograms depict the $\lambda$ values estimated from Equation 6 for the JASPAR non-redundant core motifs in \emph{S. cerevisiae} (A), \emph{D. melanogaster} (B) and available vertebrates (C) \cite{mathelier2013jaspar}.}

\end{center}

\end{figure}

\begin{figure}[h!]

\begin{center}

\includegraphics[width = 10cm]{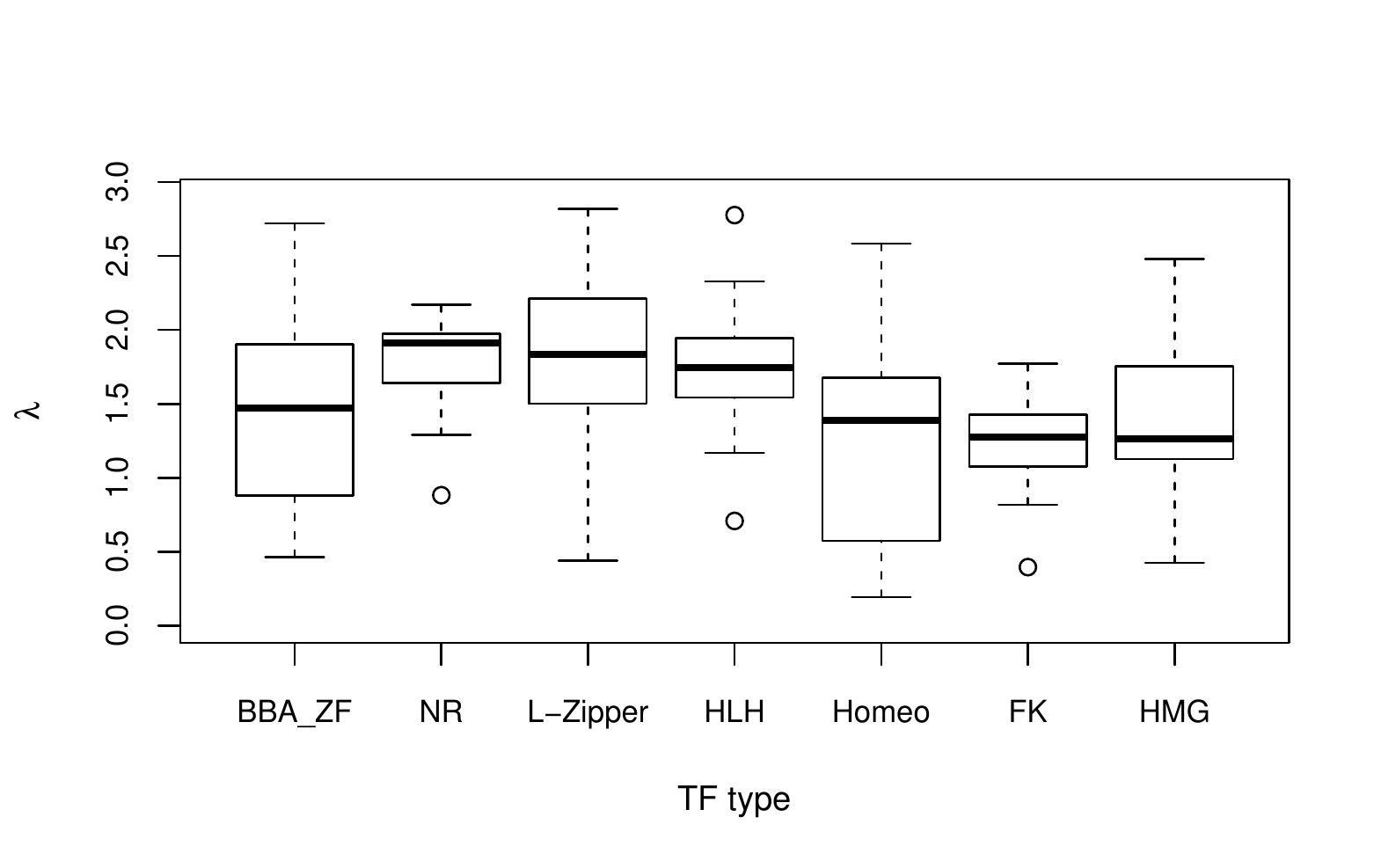}

\caption{\emph{$\lambda$ distribution comparison across major TF families.} BBA-ZF represents the $\lambda$ distribution for $\beta$-$\beta$-$\alpha$ zinc-finger family; NR is zinc-finger nuclear receptor family; L-zipper stands for the basic leucine-zipper family; HLH is helix-loop-helix family; Homeo is homeobox family; FK is fork-head family and HMG is high mobility group family. For each group, $\lambda$ was calculated by Equation 6.}

\end{center}

\end{figure}

\begin{figure}[h!]

\centering

\includegraphics[width=11cm,height=8cm]{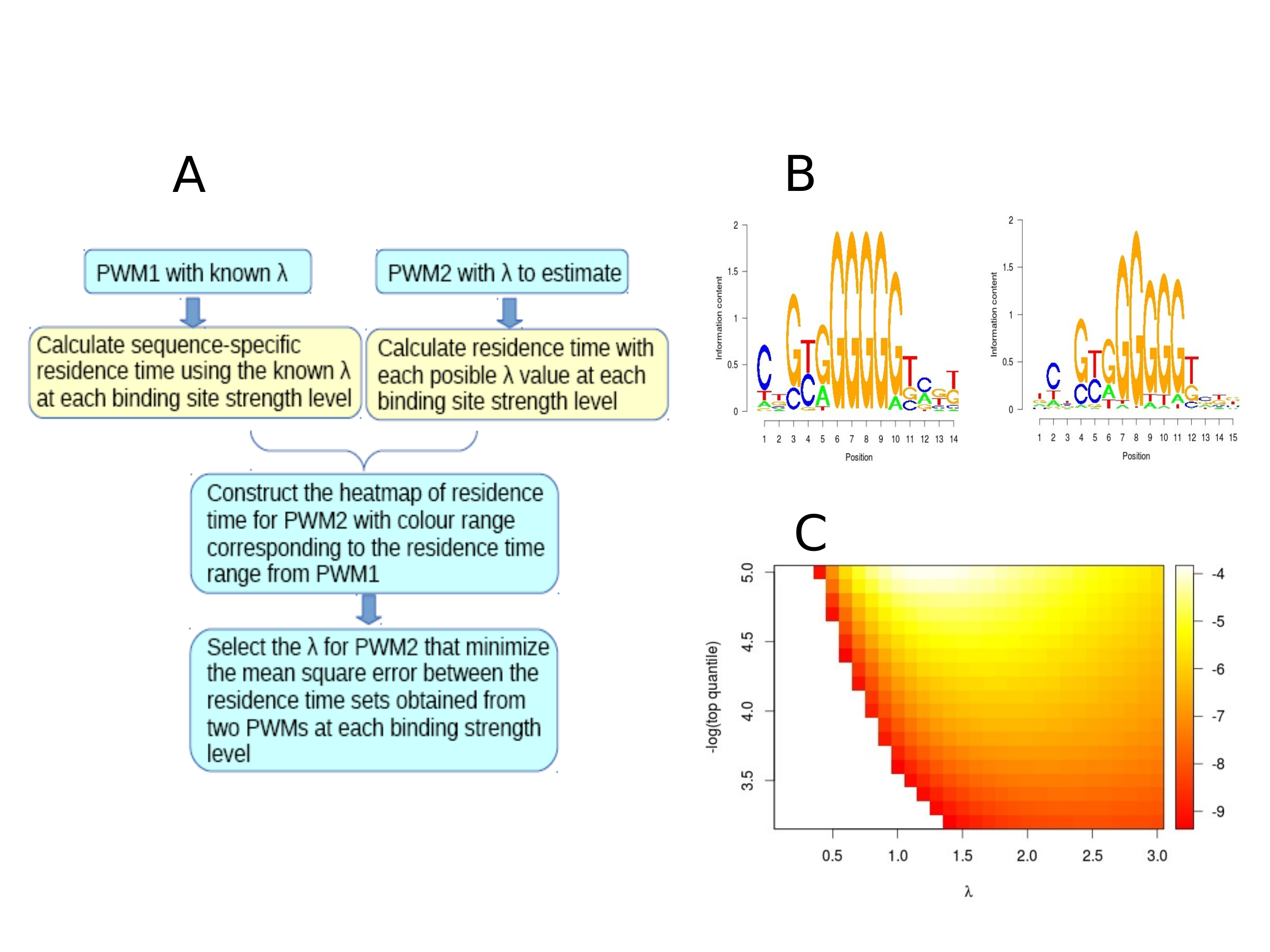}

\centering

\includegraphics[width = 11cm,height=4cm]{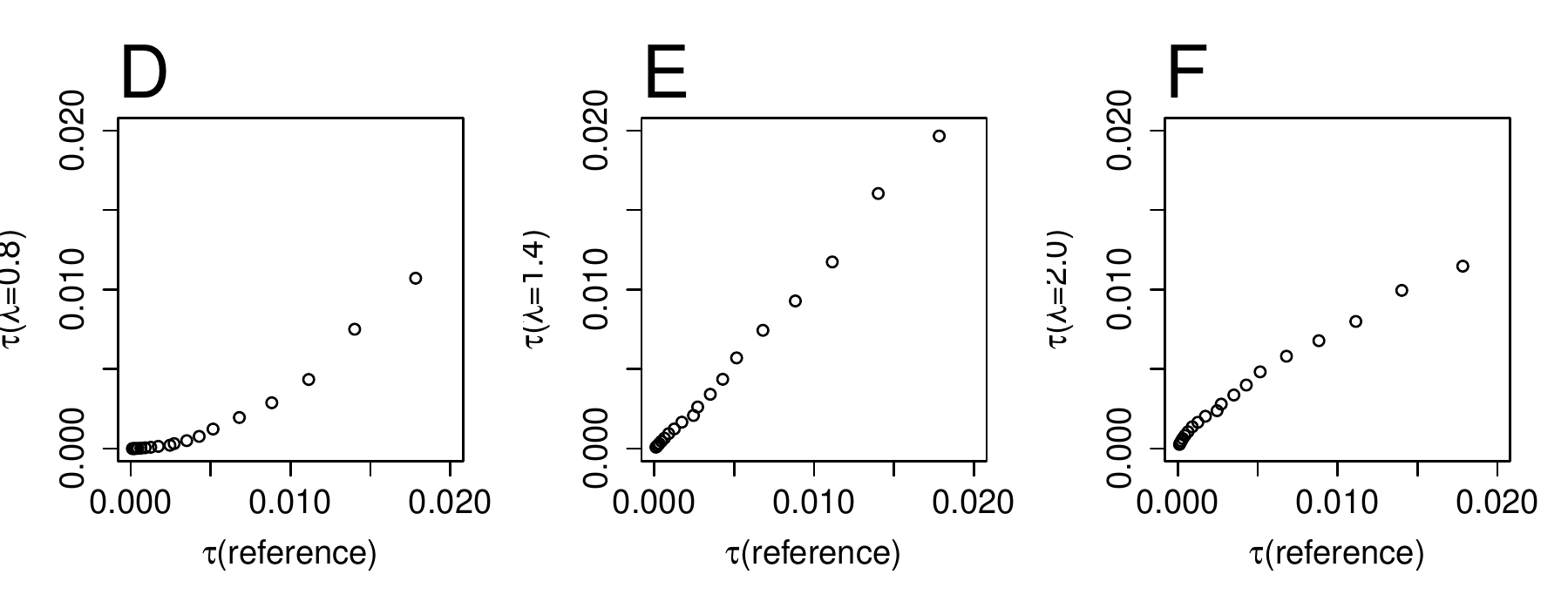}

\caption{\emph{Conversion of $\lambda$ between two PWM matrices for the lmd transcription factor.} The flow chart shows the procedure to obtain an optimised $\lambda$, given two different PWMs and one known and one unknown $\lambda$ (A). Subfigure B illustrates the two alternate PWMs for lmd which are available in the FLYFACTORSURVEY database \cite{zhu2011flyfactorsurvey}. Equation 6 suggests that PWM1 has a $\lambda$ of 1.6, and we are trying to find a $\lambda$ for PWM2. Subfigure C is a heatmap of the residence time distribution of PWM2 for different values of $\lambda$ and different binding site strength level. Each column of the heatmap represents a specific $\lambda$ value and each row represents a specific binding site strength level measured by the $-log_{10}$ of the corresponding top quantiles from low affinity to high affinity sites. Blank regions in the heatmap indicates $\lambda$ values will lead to residence time out of the reference scale which is an indication of unsuitable $\lambda$ values. D, E and F show the correlation of residence time between PWM1 and PWM2 using specific $\lambda$ values of 0.8, 1.4 and 2.0, respectively. The curve in subfigure E has the lowest mean square error, and so we assign PWM2 to have a $\lambda=1.4$. }

\end{figure}   
  
%%%%%%%%%%%%%%%%%%%%%%%%%%%%%%%%%%%
%%                               %%
%% Tables                        %%
%%                               %%
%%%%%%%%%%%%%%%%%%%%%%%%%%%%%%%%%%%

%% Use of \listoftables is discouraged.
%%
\section*{Tables}
\begin{table}[h!]
\caption{Maximum, minimum and the average values of $\lambda$ in 3 groups of organisms.}
      \begin{tabular}{cccc}
        \hline
           & \emph{S. cerevisiae} & \emph{D. melanogaster} &Vertebrates  \\ \hline
        maximum & 2.83 & 2.72 & 2.82\\
        minimum & 0.26 & 0.35  & 0.25\\
        mean & 1.25 & 1.40 & 1.73\\ \hline
      \end{tabular}
\end{table}

%%%%%%%%%%%%%%%%%%%%%%%%%%%%%%%%%%%
%%                               %%
%% Additional Files              %%
%%                               %%
%%%%%%%%%%%%%%%%%%%%%%%%%%%%%%%%%%%

\section*{Additional Files}
\subsection*{Supplementary Figures}
\begin{figure}
\begin{center} 
\includegraphics[width = 12cm]{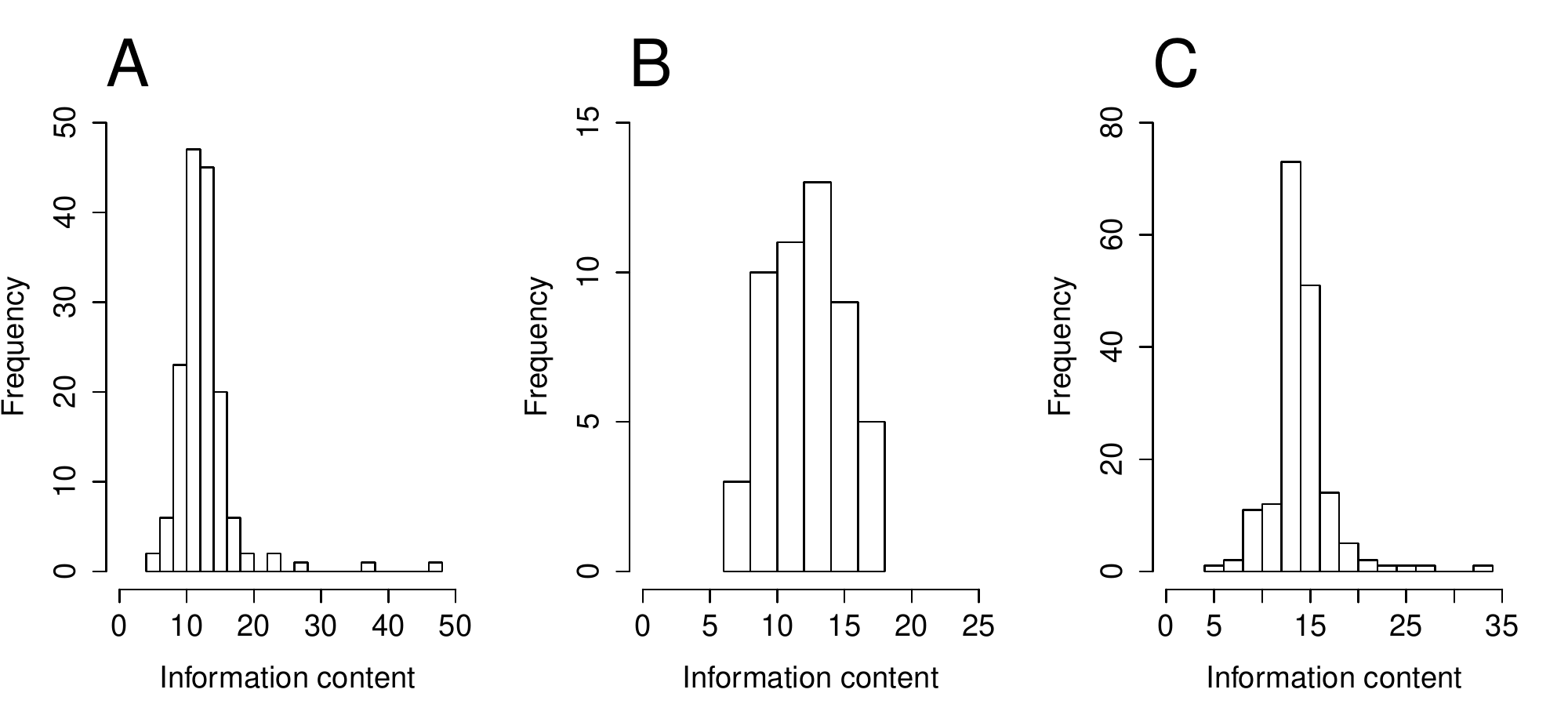} 
\caption*{Supplementary Figure 1. \emph{Information content across different organisms.} The histograms describe the information content distribution for \emph{S. cerevisiae} (A), \emph{D. melanogaster} (B) and the available vertebrates's motifs (C) from JASPAR non-redundant core motifs \cite{mathelier2013jaspar}.}
\end{center}  
\end{figure}

\begin{figure}
\begin{center} 
\includegraphics[width=10cm,height=10cm]{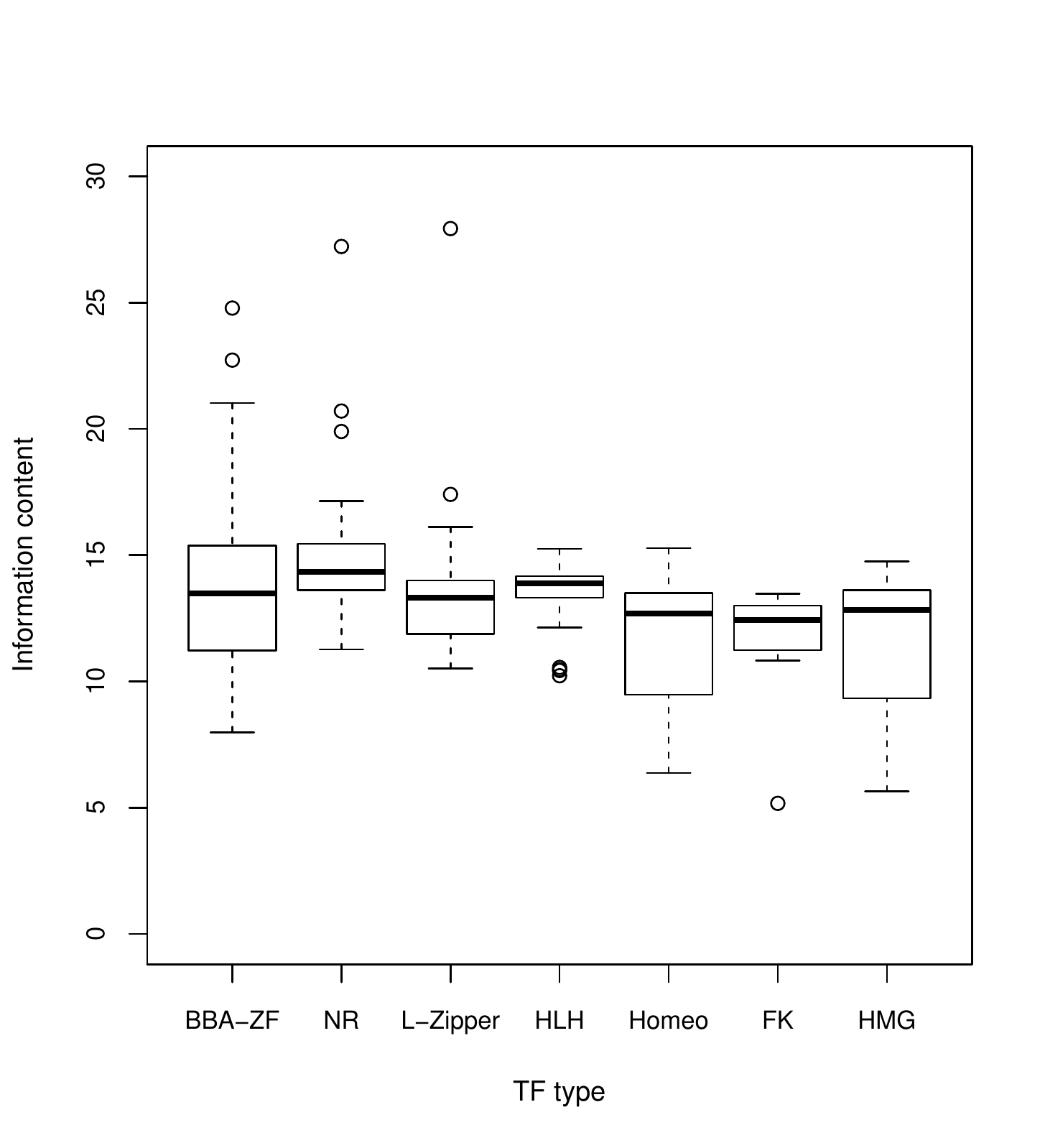} 
\caption*{ Supplementary Figure 2. \emph{Information content comparison across major TF families.} BBA-ZF represents the $\lambda$ distribution for $\beta$-$\beta$-$\alpha$ zinc-finger family; NR is zinc-finger nuclear receptor family; L-zipper stands for the basic Leucine-zipper family; HLH is helix-loop-helix family; Homeo is homeobox family; FK is forkhead family and HMG is high mobility group family.}
\end{center}  
\end{figure}

\begin{figure}
\begin{center} 
\includegraphics[width=12cm]{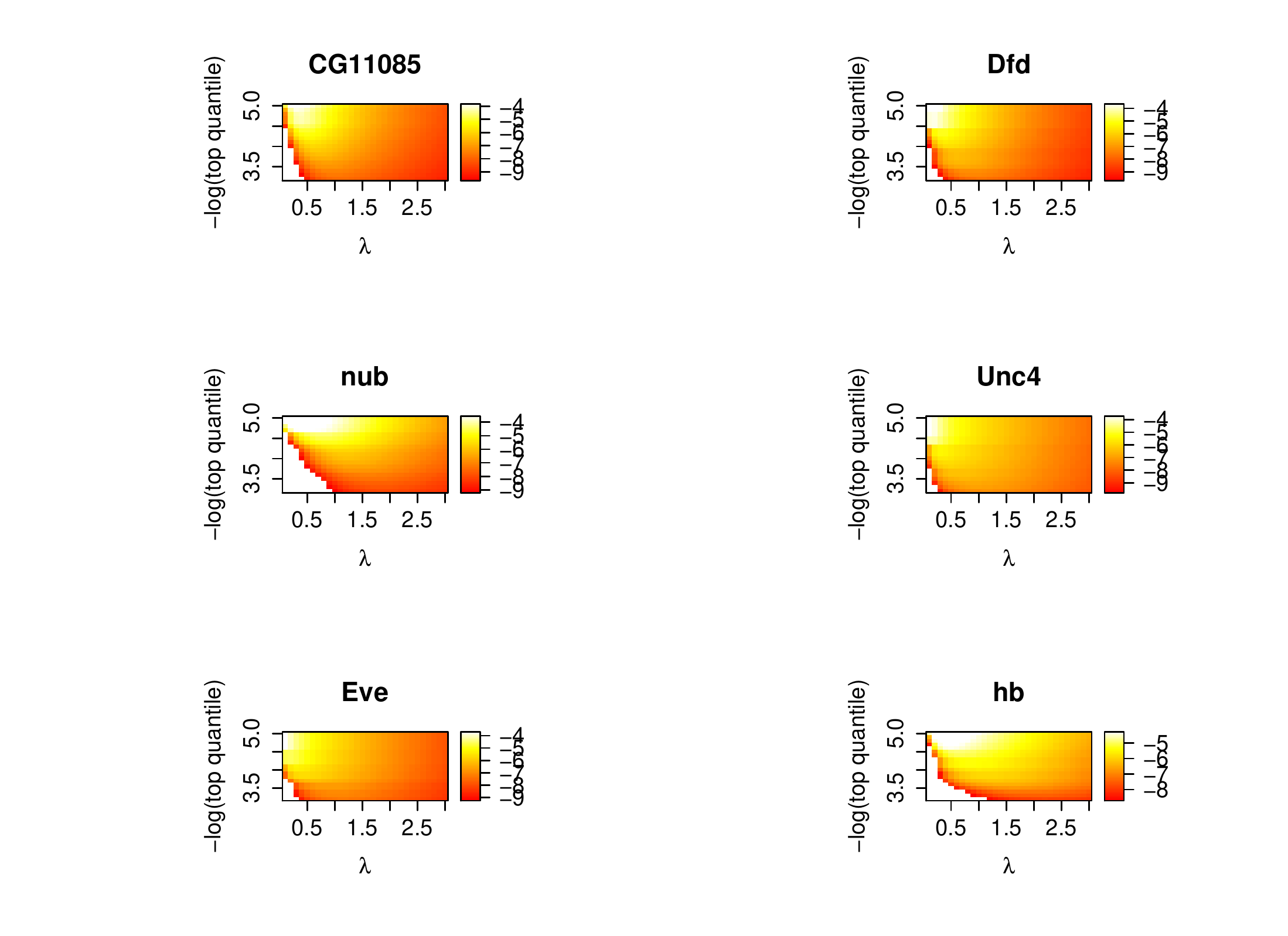} 
\caption*{Supplementary Figure 3. \emph{Heatmaps for $\lambda$ conversion between different PWMs.} These are additional examples of heatmaps of sequence-specific residence time that are used for $\lambda$ conversion between different PWM matrices of the same TF. Alternative versions of PWM matrices are from \emph{BioConductor} R package of \emph{PWMEnrich.Dmelanogaster.background} \cite{Stojnic2014PWMEnrich}. Each column of the heatmaps represents a specific $\lambda$ value and each row represents a specific binding site strength level.}

\end{center}  
\end{figure} 

\begin{figure}
\begin{center} 
\includegraphics[width = 6cm]{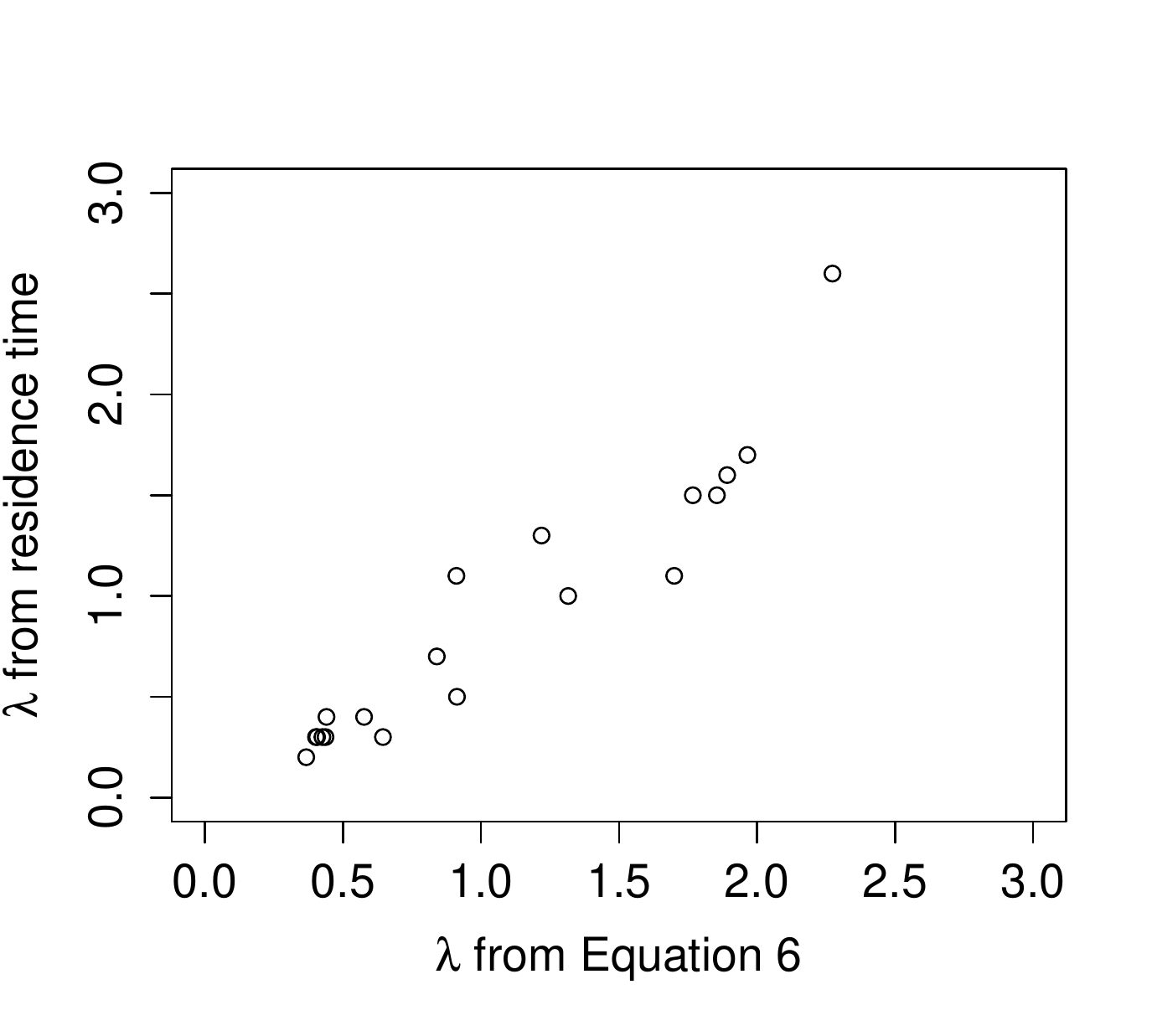} 
\caption*{Supplementary Figure 4. \emph{Consistency of $\lambda$ estimation between two methods.} This figure shows the correlation between $\lambda$ values obtained from Equation 6 and from $\lambda$ conversion using the heatmap of sequence-specific residence time. The adjusted $R^2$ is 0.88.}
\end{center}  
\end{figure} 

\begin{figure}
\begin{center} 
\includegraphics[width = 6cm]{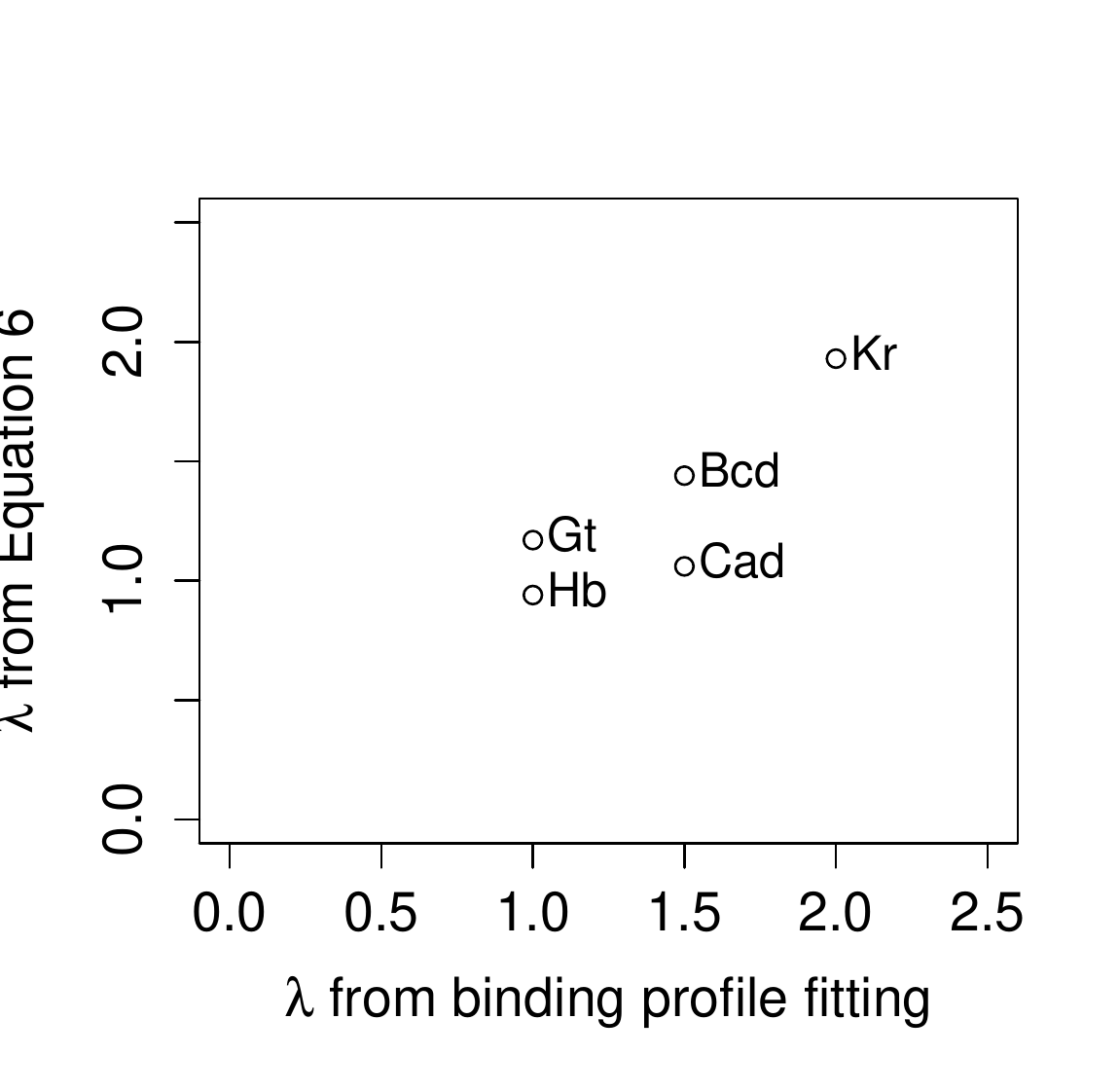} 
\caption*{ Supplementary Figure 5. \emph{ Comparison between $\lambda$ values calculated by Equation 6 and fitting ChIP-seq profile.} This figure depicts the $\lambda$ comparison between Equation 6 and previously established method using binding profile fitting for 5 \emph{D.melanogaster} TFs \cite{zabet2014estimating}. GT, HB, BCD, CAD and KR are the abbreviations for Giant, Hunchback, Bicoid, Caudal and Kruppel respectively. The adjusted $R^2$ is 0.64.}
\end{center}  
\end{figure}

\begin{figure}
\begin{center} 
\includegraphics[width = 11cm, height=6cm]{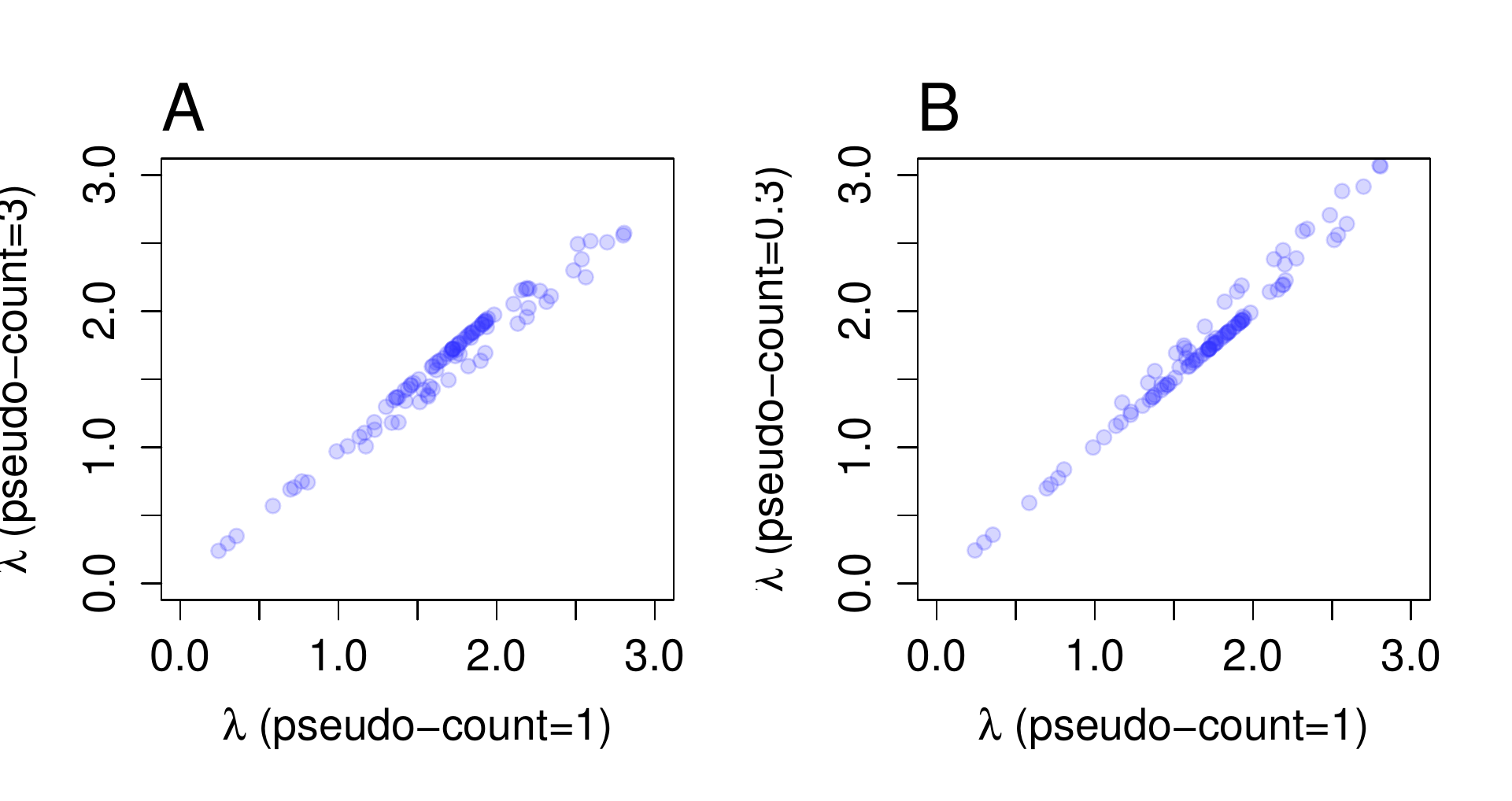} 
\caption*{ Supplementary Figure 6. \emph{ Comparison of $\lambda$ values calculated by using different pseudo-count values in PWM matrices.}  Subfigure A shows the comparison between the $\lambda$ values obtained by using PWM matrices with pseudocounts of 1 and 3 (the adjusted $R^2$ is 0.973), while subfigure B compares pseudocounts of 1 and 0.3 (the adjusted $R^2$ is 0.978). Each dot represents a TF from 100 randomly chosen vertebrate TFs in JASPAR database \cite{mathelier2013jaspar}.}
\end{center}  
\end{figure}

\end{backmatter}
\end{document}